# Learning Rarefied Gas Dynamics with Physics-Enforced Neural Networks

Ehsan Roohi[1], Ahmad Shoja-Sani[1], Bijan Goshayeshi[1], Ahmad Peyvan[2]

1. Mechanical and Industrial Engineering, University of Massachusetts Amherst,
160 Governors Dr., Amherst, MA 01003, USA
2. Division of Applied Mathematics, Brown University,
Providence, RI 02912, USA

**Abstract**

This study develops and validates neural network frameworks with physics-based constraints for surrogate modeling of rarefied gas dynamics across different levels of complexity. As a baseline, we first examine the BGK kinetic relaxation problem and show that reformulating the learning task in terms of the perturbation from the Maxwell–Boltzmann equilibrium ensures stability and accuracy. Building upon this foundation, we employ Deep Operator Networks (DeepONets) with physical constraints to address two more challenging problems. The first is the prediction of the one-dimensional structure of a standing shock wave in a rarefied polyatomic gas (pseudo-$CO_2$) at Mach 5, where the incorporation of physical constraints avoids overshoot and yields accurate predictions even for unseen viscosity ratios ($\mu_b/\mu = 50, 500$). The second is the modeling of two-dimensional rarefied hypersonic flow over a cylinder, where an ensemble of DeepONets trained on a sparse dataset ($M = 5, 7, 9$) obtained from the direct simulation Monte Carlo (DSMC) approach, generalizes successfully to both interpolation and extrapolation cases up to $M = 10$. A custom weighted loss function improves the prediction of pressure, while ensemble-based uncertainty quantification correctly identifies regions of high gradients such as shock waves. The results demonstrate that embedding physical constraints into neural operator architectures enables accurate, physically consistent, and computationally efficient surrogates, paving the way for their application to multi-dimensional high-speed rarefied flow problems.

# Introduction

The accurate simulation of high-speed rarefied gas dynamics is a cornerstone of modern aerospace engineering, underpinning the design and analysis of systems ranging from atmospheric re-entry vehicles and hypersonic transports to advanced propulsion systems (Roohi et al. 2025). Extreme physical conditions, including the presence of strong shock waves, high temperatures, and significant deviations from local thermodynamic and chemical equilibrium characterize these flow regimes (Bird 2013). For rarefied flows, where the molecular mean free path becomes comparable to the characteristic length scale, classical continuum-based models such as the Navier-Stokes-Fourier (NSF) equations are known to be inadequate (Vincenti and Kruger 1967; Roohi et al. 2025). Consequently, high-fidelity numerical methods rooted in kinetic theory, such as the Direct Simulation Monte Carlo (DSMC) method or direct solvers for the Boltzmann equation, have become the indispensable "gold standard" for achieving physically accurate predictions (Roohi et al. 2025; Bird 1994). However, the immense computational expense of these methods, which often require tracking a large number of individual particle interactions, presents a formidable barrier(Gallis et al. 2014; Moore et al. 2019). This cost becomes prohibitive for many-query applications essential to the engineering design cycle, such as uncertainty quantification (UQ), multi-objective optimization, and inverse problem-solving, thereby creating a significant bottleneck in the development of next-generation aerospace technologies.

To surmount this computational impasse, the scientific community has increasingly turned to surrogate modeling, aiming to replace expensive high-fidelity solvers with computationally efficient approximations. While traditional methods like polynomial chaos expansions and radial basis functions have seen success, their efficacy often diminishes when faced with the high-dimensional, nonlinear problems typical of fluid dynamics. The recent advent of deep learning has introduced a new class of highly expressive function approximators, offering unprecedented potential for surrogate construction (Guo et al. 2016; Xiao and Frank 2021; Miller et al. 2022; Corbetta et al. 2023; Liu et al. 2024; Tatsios et al. 2025; Chen et al. 2025; Roohi and Shoja-Sani 2025). Yet, early efforts relying on purely data-driven neural networks revealed critical limitations: a need for a large quantity of training data, a tendency to produce physically inconsistent predictions, and poor generalization to out-of-distribution scenarios (Thuerey et al. 2020). This "black-box" approach fails to leverage the physical laws governing fluid motion.

A paradigm shift occurred with the development of Physics-Informed Neural Networks (PINNs), which embed physical laws, typically in the form of partial differential equations (PDEs), directly into the network's training process via the loss function (Raissi et al. 2019). This innovation transforms the learning problem from simple curve-fitting into a constrained optimization, where the network must find a solution that not only satisfies the available data but also respects the underlying physics. This physics-informed learning acts as a powerful regularization mechanism, enabling PINNs to generalize effectively even from sparse or noisy data and ensuring that their predictions are physically plausible—a critical requirement for engineering applications (Karniadakis et al. 2021; McDevitt and Tang 2024; Fowler et al. 2024). However, a standard PINN is designed to learn the solution to a PDE for a single instance of boundary conditions and parameters. This makes it ill-suited for the parametric studies central to design exploration, as a new network would need to be trained for each point in the design space, leading to prohibitive computational costs. Recent works have explored fine-tuning strategies to accelerate training and improve convergence in such cases (Takao and Ii 2025).

The next logical evolution is to move from learning a single solution to learning the solution operator itself—the mapping from a set of input parameters or functions to the corresponding solution function. The Deep Operator Network (DeepONet) architecture has emerged as a powerful and theoretically grounded framework for this task (Lu et al. 2021). A DeepONet employs a dual-network structure: a "Branch" network processes the input parameters (e.g., Mach number, viscosity ratio), while a "Trunk" network processes the domain coordinates (e.g., spatial location). Their outputs are combined to approximate the solution field. By integrating the principles of PINNs with the operator learning capability of DeepONets, the DeepONet with physics-enforced constraints has been established as a state-of-the-art framework for creating parametric surrogates that are both data-efficient and physically consistent (Wang et al. 2021). This composite architecture learns the physical operator that governs an entire family of solutions, making it an ideal candidate for tackling complex, multi-query problems in fluid dynamics (Cai et al. 2021; Sun et al. 2020).

This paper leverages the DeepONet and other frameworks with some physical constraints to address three distinct and highly challenging problems in gas dynamics, chosen specifically to stress-test the methodology across different axes of complexity.

The first challenge is the solution of the kinetic relaxation problem governed by the Bhatnagar–Gross–Krook (BGK) model. This case serves as a validation of stability and accuracy for physics-informed learning, where we demonstrate that reformulating the learning task in terms of the perturbation from the Maxwell–Boltzmann equilibrium distribution allows the network to avoid trivial non-physical solutions.

The second challenge is the prediction of the one-dimensional structure of a standing shock wave in a rarefied polyatomic gas (pseudo-$CO_2$). This problem is a test of physical fidelity, as the large bulk viscosity and slow relaxation of internal energy modes in such gases give rise to complex, non-equilibrium shock profiles that defy classical theories (Taniguchi et al. 2014). Capturing the transition between symmetric, asymmetric, and double-layer shock structures as a function of physical parameters is a formidable task that probes the model's ability to learn intricate, non-equilibrium physics (Kosuge and Aoki 2018).

The third challenge is the prediction of two-dimensional rarefied hypersonic flow over a cylinder. This problem is a test of data efficiency and parametric generalization in a geometrically complex setting. Here, the objective is to train a surrogate on an extremely sparse dataset of only three Mach numbers (M=5, 7, 9) and demonstrate its ability to accurately interpolate and extrapolate the entire flow field across a wide, continuous range of unseen Mach numbers (M=5.5 to M=10).

The approach presented in this paper leverages neural operators for complex physical systems. Recent studies have successfully applied DeepONet architectures to a wide array of challenging problems, including general compressible flows (Zheng et al. 2025), spectral analysis of explosion separation events (Chen et al. 2025), and non-equilibrium chemistry in hypersonic shocks (Zanardi et al. 2023). Concurrently, significant efforts are focused on enhancing the data efficiency and interpretability of these models. For example, Barwey et al. (Barwey et al. 2025) employed multiscale graph neural networks for mesh-based super-resolution of fluid flows, highlighting the growing variety of surrogate architectures in this domain. Fusion-DeepONet has been developed for geometry-dependent hypersonic flows (Peyvan et al. 2025), while RiemannONets aim to create interpretable operators for fundamental Riemann problems (Peyvan et al. 2024). These advancements, alongside related methodologies like differentiable programming for both continuum and rarefied flows (Xiao 2025), underscore a clear trajectory in computational science: the fusion of deep learning with physical principles.

In brief, the BGK relaxation problem tests the model's ability to solve fundamental kinetic relaxation dynamics and highlights the contrast between forward and inverse formulations. The 1D shock problem tests the model's ability to capture complex, non-equilibrium physics from parametric data (viscosity ratio). The 2D cylinder problem tests its ability to generalize across a geometric domain from extremely sparse parametric data.

This work makes several key contributions to the field of scientific machine learning for aerospace applications. First, we develop and validate a stable and accurate PINN-based framework for the BGK relaxation problem, demonstrating its ability to recover the correct equilibrium distribution. Second, we construct a DeepONet with physics-enforced constraints surrogate for complex 1D polyatomic shock structures, demonstrating its ability to accurately predict non-equilibrium profiles for unseen physical parameters with high fidelity. Third, we develop a robust DeepONet surrogate for 2D hypersonic cylinder flow that generalizes remarkably well from a severely sparse training dataset, showcasing its potential for rapid design space exploration. Finally, we present a methodological investigation into model refinement, demonstrating how an ensemble architecture can provide reliable uncertainty quantification and how a custom weighted loss function can be strategically employed to improve the accuracy of specific, hard-to-learn physical quantities like pressure. The results confirm that the model's uncertainty is highest in regions of high physical gradients, such as the shock wave. This study collectively demonstrates that PINNs and DeepONets with physics-enforced constraints are powerful, reliable, and versatile tools for creating efficient surrogate models for complex, multi-dimensional problems in high-speed aerodynamics.

The remainder of this paper is organized as follows. Section 2 introduces the BGK relaxation problem and its PINN formulation. Section 3 details the physical formulation for the polyatomic gas shock wave and its problem setup. It describes our physics-enforced constraints DeepONet architecture, physics-informed training methodology, and uncertainty quantification strategy. Following this, we present and discuss the quantitative and qualitative results for the shock structure. Section 4 describes the neural network for the hypersonic cylinder problem, followed by a discussion of the results. Section 5 provides concluding remarks.

# BGK Relaxation problem

## Problem Formulation and the Perturbation Ansatz

We consider monoatomic argon with molecular mass $m = 6.6335 \times 10^{-26}$ kg at equilibrium temperature $T_{eq} = 273$ K and mass density $\rho_{eq} = 1.78$ kg/m$^3$, giving a number density $n_{eq} = \rho_{eq}/m \approx 2.68 \times 10^{25}$ m$^{-3}$. The characteristic (thermal) speed is $v_c = \sqrt{2k_B T_{eq}/m} \approx 3.37 \times 10^2$ m/s and the characteristic relaxation time is $\tau_{rel} = 10^{-10}$ s. We non-dimensionalize via $\hat{v} = v/v_c$ and $\hat{t} = t/\tau_{rel}$, and use the standard Maxwell–Boltzmann equilibrium for the speed distribution.

The relaxation of a gas towards equilibrium is described by the BGK model. The governing equation is:

$$\frac{\partial P}{\partial t} = \frac{w(t)}{\tau}(P_M - P),$$

where $P(v,t)$ is the speed distribution function, $P_M(v)$ is the known equilibrium Maxwell-Boltzmann distribution, and $w(t)$ is a time-dependent weight. For the forward problem, the characteristic relaxation time, $\tau$, is treated as a known physical parameter.

A direct attempt to learn $P(v,t)$ using a neural network failed due to convergence to trivial non-physical solutions. To overcome this, we reformulate the problem by tasking the network to learn only the deviation from equilibrium, $\Phi(v,t)$, via the ansatz:

$$P(v,t) = P_M(v) \cdot (1 + \Phi(v,t))$$

The Maxwell-Boltzmann distribution $P_M(v)$ is given by:

$$P_M(v) = 4\pi \left(\frac{m}{2\pi k_B T}\right)^{3/2} v^2 \exp\left(-\frac{mv^2}{2k_B T}\right)$$

Substituting the ansatz into the BGK equation yields a stable decay equation for the perturbation:

$$\frac{\partial \Phi}{\partial t} + \frac{w(t)}{\tau}\Phi = 0$$

The initial condition and the law of particle conservation are also reformulated in terms of $\Phi$:

$$\Phi(v, t=0) = \frac{P_{\text{initial}}(v)}{P_M(v)} - 1$$

$$\int_0^\infty P_M(v)\Phi(v,t)\,dv = 0$$

This transforms a difficult regression task into a much simpler problem of learning a function that decays towards zero.

## Network Architecture and Training Strategy

The neural network approximating the perturbation function, $\Phi_{NN}(t, v; \boldsymbol{\theta})$, is a fully-connected feed-forward network with 2 input neurons $(t, v)$, 5 hidden layers of 60 neurons each, and a single output neuron for $\Phi$. A two-phase hybrid training approach was employed, leveraging the complementary strengths of the Adam and L-BFGS optimizers. The loss function includes the PDE and initial condition residuals, as well as a soft constraint enforcing the particle conservation law. The schematic of the employed network is shown in Fig. 1.

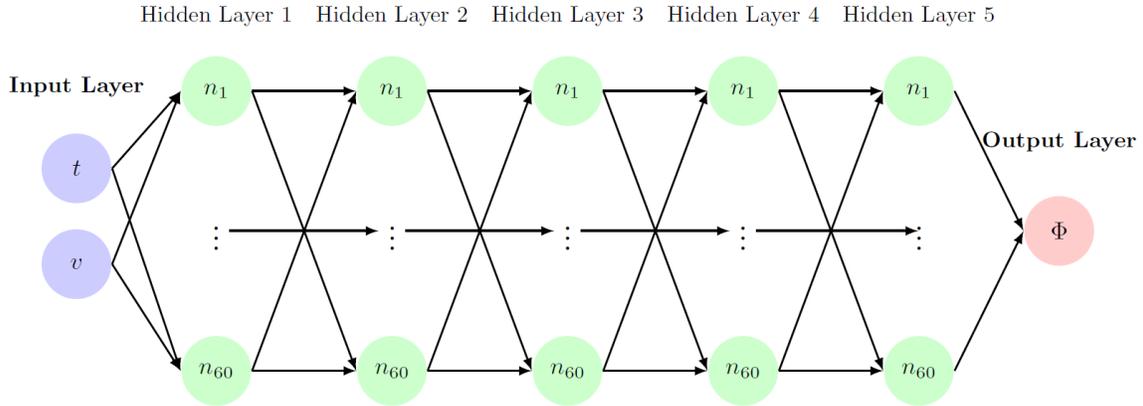

*Fig. 1: Schematic of the neural network architecture employed in this work. The network is a fully-connected, feed-forward architecture consisting of 2 input neurons (t, v), 5 hidden layers with 60 neurons each, and 1 output neuron for predicting the perturbation function ($\Phi$).*

### Phase 1: Global Exploration with Adam

The first phase of training uses the Adam (Adaptive Moment Estimation) optimizer. Adam is a first-order, gradient-based algorithm that is well-suited for the initial, exploratory phase of training. Let $\boldsymbol{\theta}$ be the parameters of the neural network and $\mathcal{L}(\boldsymbol{\theta})$ be the total loss function. At each step $t$, Adam performs the following updates:

1. Compute the gradient: $\mathbf{g}_t = \nabla_{\boldsymbol{\theta}} \mathcal{L}(\boldsymbol{\theta}_{t-1})$
2. Update the first moment (momentum) estimate: $\mathbf{m}_t = \beta_1 \mathbf{m}_{t-1} + (1 - \beta_1)\mathbf{g}_t$
3. Update the second moment (uncentered variance) estimate: $\mathbf{v}_t = \beta_2 \mathbf{v}_{t-1} + (1 - \beta_2)\mathbf{g}_t^2$
4. Compute bias-corrected moment estimates: $\hat{\mathbf{m}}_t = \frac{\mathbf{m}_t}{1-\beta_1^t}$ and $\hat{\mathbf{v}}_t = \frac{\mathbf{v}_t}{1-\beta_2^t}$
5. Update the parameters: $\boldsymbol{\theta}_t = \boldsymbol{\theta}_{t-1} - \eta \frac{\hat{\mathbf{m}}_t}{\sqrt{\hat{\mathbf{v}}_t} + \epsilon}$

Here, $\eta$ is the learning rate, $\beta_1$ and $\beta_2$ are exponential decay rates for the moment estimates, and $\epsilon$ is a small constant for numerical stability.

### Phase 2: Local Refinement with L-BFGS

After the initial training with Adam, optimization is switched to the L-BFGS (Limited-memory Broyden–Fletcher–Goldfarb–Shanno) algorithm. L-BFGS is a quasi-Newton method that approximates the inverse Hessian matrix to perform a more informed, second-order update. The general update rule for a Newton-like method is:

$$\boldsymbol{\theta}_{k+1} = \boldsymbol{\theta}_k - \alpha_k \mathbf{H}_k^{-1} \mathbf{g}_k$$

where $\mathbf{g}_k$ is the gradient, $\mathbf{H}_k^{-1}$ is the inverse of the Hessian matrix at step $k$, and $\alpha_k$ is the step size found via a line search. This makes it highly efficient for the final fine-tuning stage of training.

### Forward Problem Solution

The primary goal of this work was to develop a stable PINN solver for the BGK relaxation problem. For this forward problem, the theoretically-derived non-dimensional relaxation time ($\hat{\tau}^{-1} \approx 0.467$) was used as a fixed input parameter. The results, shown in Figure 2, confirm the success of this approach. The training loss converges smoothly, and the final predicted speed distribution shows a visually perfect match with the theoretical Maxwell-Boltzmann equilibrium state. This demonstrates that the perturbation ansatz provides a robust and accurate framework for solving the BGK relaxation equation when the physical parameters are known.

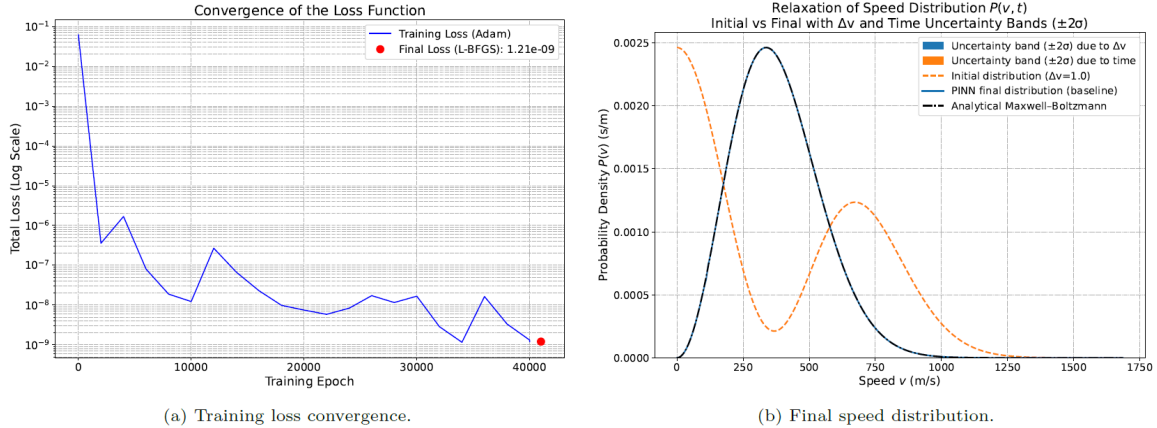

(a) Training loss convergence.

(b) Final speed distribution.

*Figure 2: Results for the forward problem, showing stable convergence (a) and an accurate final distribution matching the theoretical equilibrium (b).*

# Shock Structure in Polyatomic Rarefied Gas

## Problem Statement

Shock waves have a different structure in rarefied polyatomic gases than in monatomic gases. This is because polyatomic molecules possess internal rotational and vibrational modes that can store energy, which single-atom molecules lack (Griffith and Kenny 1957; Zel'dovich and Raizer 2002). Shock waves in rarefied polyatomic gases have two distinct characteristics: thickness and a speed-dependent shape. Regarding thickness, a major difference exists compared to monatomic gases. While a shock wave in a monatomic gas is extremely thin, i.e., on the order of one mean free path, in a polyatomic gas it is much thicker—often several orders of magnitude larger and potentially centimeters long. This increased thickness occurs because the internal rotational and vibrational modes of polyatomic molecules require a longer time to "relax" and transfer their energy, which effectively smears the shock wave out over a larger distance. The second characteristic is the evolution of the shock wave's profile with its speed, measured by the Mach number. As the Mach number increases, the profile changes from a nearly symmetric shape (Type A) at low Mach numbers, to an asymmetric profile (Type B) at medium Mach numbers, and finally to a complex structure composed of distinct thin and thick layers (Type C) at high Mach numbers. The structure of a shock wave in a polyatomic gas is significantly influenced by the ratio of bulk viscosity ($\mu_b$) to shear viscosity ($\mu$). For gases like $CO_2$, this ratio can be very large, on the order of 1000 (Kosuge and Aoki 2018). This large bulk viscosity leads to a broadening of the shock wave and can result in complex shock profiles. Depending on the upstream Mach number, these profiles can be classified into three types, as shown in Figure 3 (Kosuge and Aoki 2018; Taniguchi et al. 2014).

For very weak shock waves (Mach number close to 1), the profiles of density, velocity, and temperature are nearly symmetric (**Type A**). As the Mach number increases, the profiles become asymmetric, exhibiting a sharp corner-like feature on the upstream side (**Type B**). At even higher Mach numbers, a **Type C** double-layer structure emerges. This consists of a thin upstream layer

Δ with a steep change in physical quantities, followed by a much thicker downstream layer Ψ where the variables slowly approach their equilibrium values.

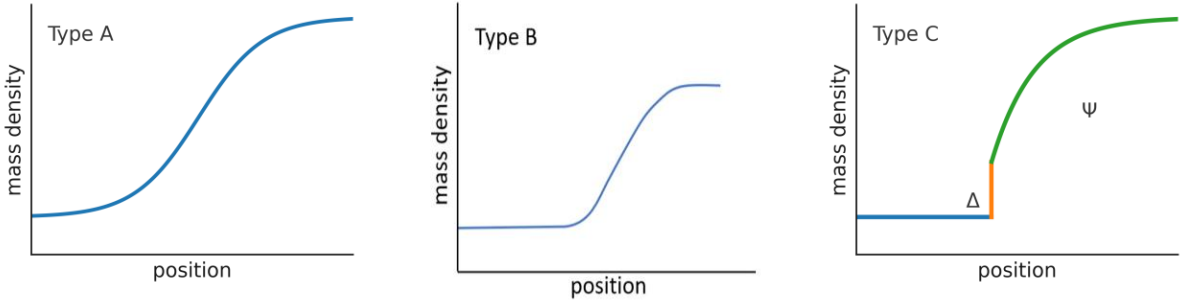

*Type A: Symmetric profile.*    *Type B: Asymmetric profile.*    *Type C: Double-layer structure.*

*Fig. 3: The three types of shock wave structures in a rarefied polyatomic gas. For very weak shock waves (a), the profile is nearly symmetric. As the Mach number increases (b), it becomes asymmetric with a sharp upstream corner. At higher Mach numbers (c), a double-layer structure emerges.*

## Problem Formulation

We consider a stationary plane shock wave in a polyatomic gas. The gas is in an equilibrium state at upstream infinity ($x_1 \to -\infty$) with density $\rho_-$, velocity $v_-$, and temperature $T_-$. At downstream infinity ($x_1 \to \infty$), the gas is in another equilibrium state with density $\rho_+$, velocity $v_+$, and temperature $T_+$. The behavior of the gas is described by the ellipsoidal statistical (ES) model of the Boltzmann equation for a polyatomic gas in (Kosuge and Aoki 2018). The ES data from this reference was used to train the DeepONet with physics-enforced constraints used in the current study. The problem is treated as one-dimensional, with all physical quantities depending only on the spatial coordinate $x_1$.

To analyze the problem, we use a set of dimensionless variables. The data contains values for the following dimensionless quantities (Kosuge and Aoki 2018):

**Position ($x_1$):** The spatial coordinate normalized by a reference length scale. **Density ($\hat{\rho}$):** The local density normalized by the upstream density, $\hat{\rho} = \frac{\rho}{\rho_-}$. **Velocity ($\hat{v}_1$):** The local velocity in the direction of the shock wave, normalized by the upstream velocity, $\hat{v}_1 = \frac{v_1}{v_-}$. **Temperature ($\hat{T}$):** The local temperature normalized by the upstream temperature, $\hat{T} = \frac{T}{T_-}$. **Pressure ($\hat{p}$):** The local pressure normalized by the upstream pressure, $\hat{p} = \frac{p}{p_-}$. **Translational Temperature ($\hat{T}_{tr}$):** The temperature associated with the translational motion of the gas molecules, normalized by the upstream temperature. **Internal Temperature ($\hat{T}_{int}$):** The temperature associated with the internal energy modes of the gas molecules, normalized by the upstream temperature. **Stress Components ($\hat{p}_{11} - \hat{p}, \hat{p}_{22} - \hat{p}$):** These represent the deviatoric parts of the stress tensor, which are measures of the departure from isotropic pressure. **Heat Flux ($-\hat{q}_1$):** The component of the heat flux vector in the direction of the shock wave.

These dimensionless variables are used to train the neural network at various viscosity ratios.

## Neural Network for Density Prediction

The numerical simulation of shock-wave structures, especially for gases with large bulk viscosity, can be computationally intensive. This is due to the need for a very large computational domain to capture the thick downstream layer of a Type C profile, considered in this study, along with a fine grid to resolve the thin upstream layer (Kosuge and Aoki 2018).

This is where the application of a neural network becomes particularly valuable. By training a neural network on a set of numerical solutions for various ratios of bulk viscosity, it is possible to create a model that can accurately predict the shock structure for other values of this ratio. This work in successfully predicting the shock structure for a bulk viscosity ratio of 50 demonstrates the potential of this approach.

The key advantages of using a neural network in this context are: Computational Efficiency, Interpolation, and Prediction and Surrogate Modeling. The core of the model is a DeepONet, which is specifically designed to learn operators between function spaces. It consists of two main sub-networks: a Branch Network and a Trunk Network.

Branch Network processes the input function, which in this case is represented by the scalar parameter defining the physical conditions, i.e., the viscosity ratio $u = (\mu_b/\mu)$. It encodes this parameter into a latent space representation. Trunk Network processes the spatial coordinate of the output function, $y = (x_1)$. It generates a set of basis functions that can be used to construct the solution profile.

The outputs of these two networks are then combined via a dot product to approximate the final solution $G(u)(y) \approx \rho(x_1)$. A schematic of this architecture is shown in Figure 4.

## Physics-Informed Constraints

To ensure the physical realism of the predictions, three critical constraints were embedded into the model's loss function during training:

1. **Bounded Output:** The density $\rho$ is a normalized quantity. To enforce that its value always remains within the physical bounds [0, 1], the final activation function of the network is a Sigmoid function. This is combined with a 'MinMaxScaler' on the training data.

2. **Monotonicity Constraint:** Across a shock wave, the density must be a non-decreasing function of the spatial coordinate $x_1$. This physical law is enforced by adding a penalty term to the loss function for any instance where the predicted gradient is negative:

$$\mathcal{L}_{mono} = \mathbb{E}\left[\text{ReLU}\left(-\frac{\partial \rho}{\partial x_1}\right)^2\right]$$

The symbol $\mathbb{E}$ denotes the expectation operator (statistical average). The function $\text{ReLU}(z)$ is the Rectified Linear Unit, widely used in machine learning, and is defined as

$$\text{ReLU}(z) = \max(0, z),$$

that is, it returns $z$ when $z > 0$, and $0$ otherwise. This forces the model to learn solutions where $\frac{\partial \rho}{\partial x_1} \geq 0$ everywhere.

3. **Boundary Conditions:** Far from the shock front (at $x_1 \to \pm\infty$), the flow variables must settle to constant free-stream values, meaning their gradients must approach zero. This is enforced by penalizing non-zero gradients in the far-field regions:

$$\mathcal{L}_{boundary} = \mathbb{E}\left[\left(\frac{\partial \rho}{\partial x_1}\bigg|_{x_1 \to \pm\infty}\right)^2\right]$$

The total loss function is a weighted sum of the standard data-driven Mean Squared Error (MSE) loss and these physics-based penalty terms. An annealing schedule is used to gradually increase the weight of the physics losses, allowing the model to first fit the data and then refine its predictions to be consistent with physical laws. Figure 4 shows the employed network. Unlike a vanilla multilayer perceptron (MLP), which is trained solely on data without any guarantee of physical consistency, the proposed Physics-influenced DeepONet architecture explicitly incorporates domain knowledge through physical constraints. As shown in Fig. 4, the model consists of two subnetworks: a branch net that encodes the input parameter $\mu_b/\mu$, and a trunk net that encodes the spatial coordinate $x_1$. Their outputs are combined to predict the density profile $\rho(x_1)$. During training, in addition to minimizing the data loss, the model enforces three physical constraints: (i) bounded output via sigmoid activation, ensuring $\rho \in [0,1]$; (ii) monotonicity, $\partial \rho / \partial x_1 \geq 0$; and (iii) boundary conditions, $\partial \rho / \partial x_1|_{x \to \pm\infty} = 0$. These additional constraints prevent nonphysical solutions and significantly improve generalization compared to a vanilla network.

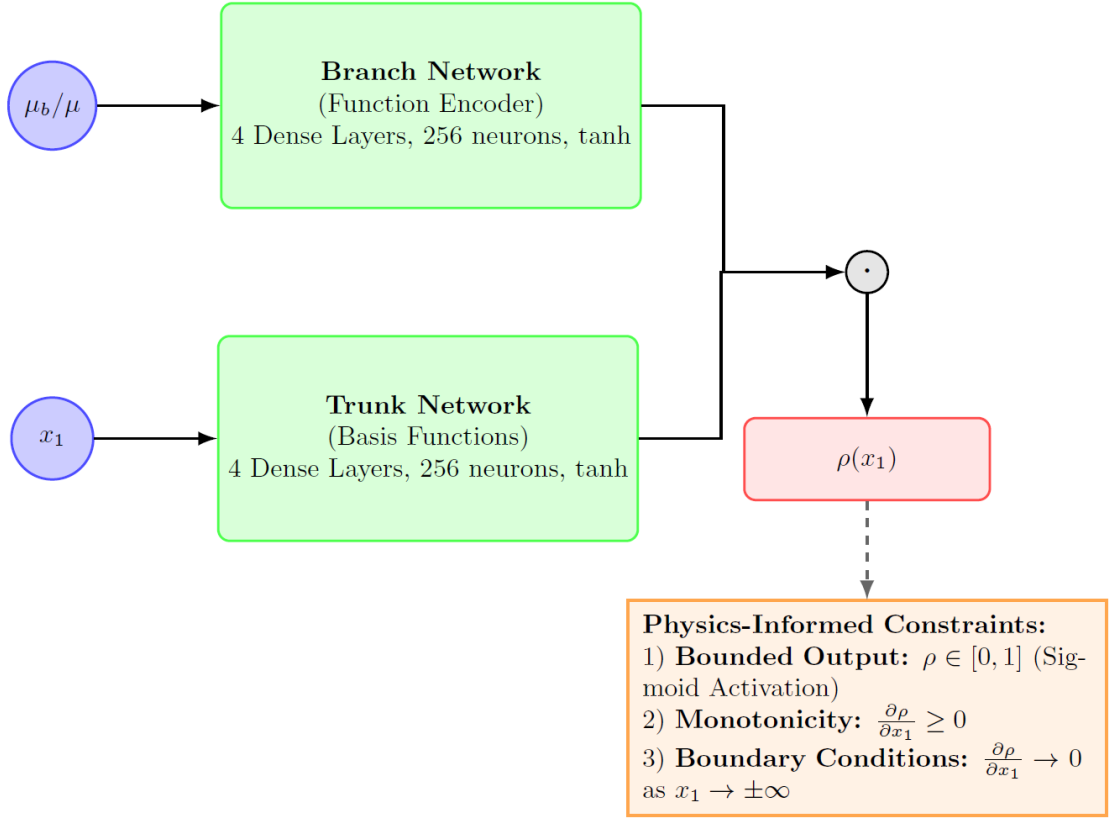

*Fig. 4: Schematic of the Physics-Informed DeepONet for Density Prediction. The Branch Network encodes the scalar input parameter $\mu_b/\mu$, while the Trunk Network encodes the spatial coordinate $x_1$. Their outputs are combined via a dot product to predict the density profile $\rho(x_1)$. Physics-informed constraints, including boundedness, monotonicity, and far-field boundary conditions, are embedded into the model's loss function during training.*

## Results and Discussions

The trained model demonstrated remarkable accuracy, successfully predicting the density profile for a test case with a viscosity ratio not seen during training. Figure 5 illustrates the training history of the proposed bounded monotonic network. The training and validation losses (measured in mean squared error, MSE) both decrease steadily with epochs, demonstrating convergence of the data-driven objective. The validation loss remains close to the training loss, indicating the absence of significant overfitting.

In addition to the data loss, physics-based penalties are also monitored. The monotonicity loss decreases by several orders of magnitude, showing that the network successfully enforces the physical constraint $\partial \rho / \partial x_1 \geq 0$. The boundary gradient loss remains at a very small scale ($10^{-8}$ to $10^{-7}$), confirming that the boundary condition $\partial \rho / \partial x_1|_{x \to \pm\infty} = 0$ is satisfied throughout training. The annealed physics weight (purple curve) increases gradually, ensuring a balanced contribution of the physical constraints to the total loss.

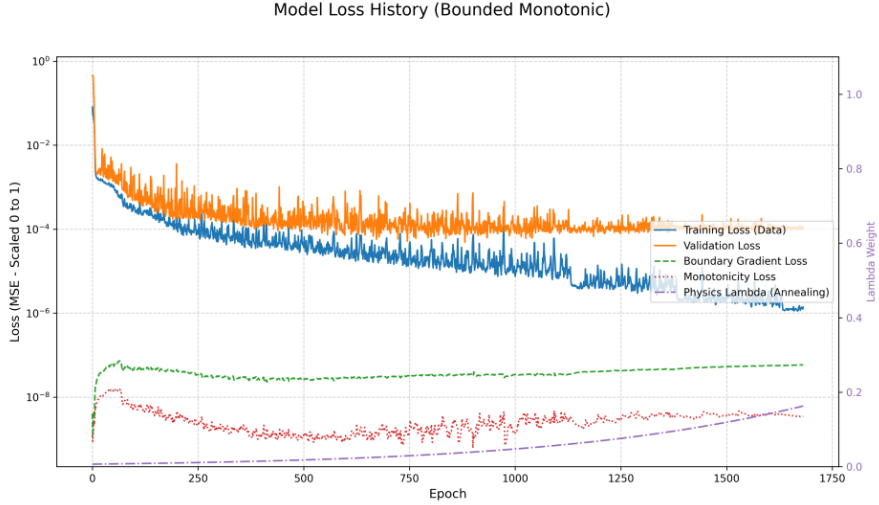

*Fig. 5: Training and validation loss history over epochs. This plot shows the convergence of the data loss (MSE) as well as the physics-based losses for monotonicity and boundary gradients.*

Figures 6 and 7 demonstrate the predictive capability of the bounded monotonic DeepONet architecture across both training and unseen cases at a Mach number of 5 at various viscosity ratios. $\gamma$ is expressed in terms of the internal degrees of freedom $\delta$ of a molecule as

$$\gamma = \frac{\delta + 5}{\delta + 3},$$

where $\delta$ can be any positive real number (not restricted to an integer). The presented results are for

$$\delta = 4 \quad (\gamma = 9/7), \qquad Pr = 0.761.$$

In Figure 8, the model fit is shown for a range of training cases with different viscosity ratios $\mu_b/\mu \in \{10, 20, 100, 200, 500, 1000, \infty\}$. Across all training scenarios, the predicted density profiles (dashed red lines) overlap almost perfectly with the true simulation data (solid blue lines). This agreement confirms that the network successfully learns the mapping between the input parameter $\mu_b/\mu$ and the corresponding density distribution $\rho(x_1)$. Notably, the network captures both the sharp transition near $x_1 = 0$ for small viscosity ratios, as well as the smoother, extended gradients observed for larger ratios. The consistency of these fits across several orders of magnitude in $\mu_b/\mu$ highlights the robustness and expressiveness of the operator-learning framework.

Figure 9 extends this evaluation to an unseen test case with $\mu_b/\mu = 50$, which was not part of the training set. The DeepONet prediction again aligns closely with the true solution, demonstrating strong interpolation capability within the parameter space. The model is able to capture the steep rise in density and accurately reproduce the asymptotic saturation at $\rho \to 1$. This result emphasizes the generalization ability of the bounded monotonic architecture, confirming that the incorporation of physical constraints enables accurate and stable predictions even for test cases not encountered during training.

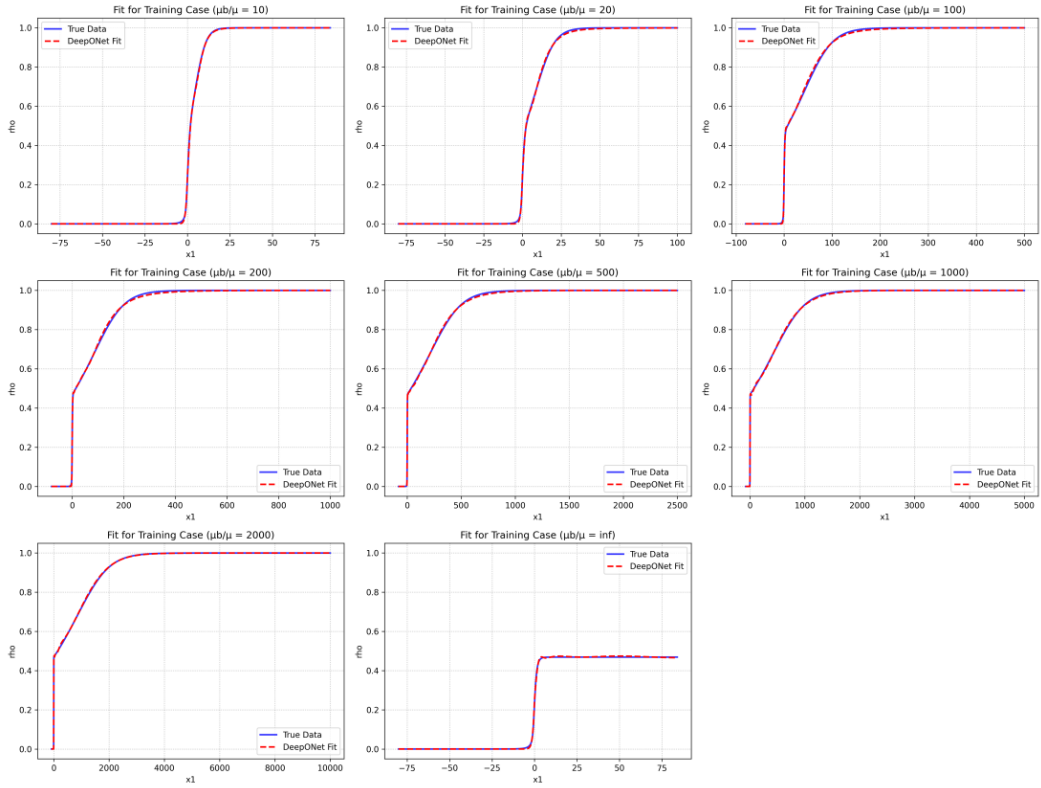

*Fig. 6: Model fit across a range of training cases with different viscosity ratios. The plots demonstrate the model's ability to accurately capture the solution profile for the entire family of parameters.*

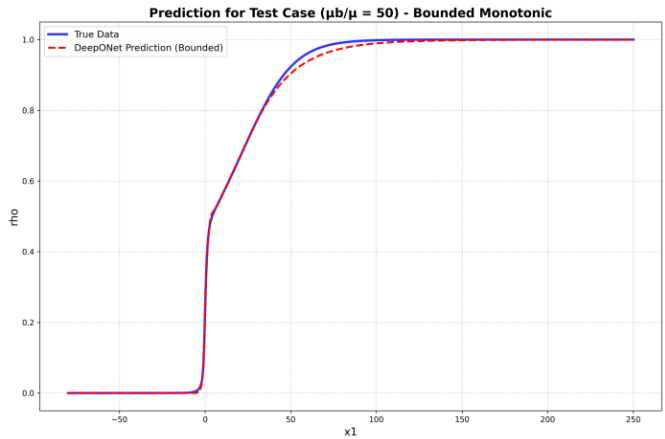

*Fig. 7: Comparison of the DeepONet prediction (dashed red line) against the true simulation data (solid blue line) for the unseen test case where $\mu_b/\mu = 50$.*

# Neural Network for General Properties Prediction

In this section, we extend the previous DNN to predict more properties of the flow field in a polyatomic shock wave. The employed neural network is a Deep Operator Network featuring a sophisticated multi-head architecture designed to accurately predict both monotonic and non-monotonic physical variables from a single model, see Fig. 8. The architecture consists of a shared core that learns a general representation of the underlying physics, which then feeds into two specialized output heads. The entire network is trained end-to-end, optimizing a composite loss function that includes both data fidelity and physical constraint terms.

As before, the shared core is a standard DeepONet, comprising two parallel multi-layer perceptrons (MLPs): a Branch Network and a Trunk Network. The Branch Net processes the input physical parameter (the viscosity ratio, $\mu_b/\mu$), while the Trunk Net processes the spatial coordinate ($x_1$). Both networks consist of two hidden layers with 512 neurons each, using the hyperbolic tangent (tanh) activation function. The outputs of these two networks, which are 192-dimensional latent vectors, are combined via a dot product to produce a shared feature representation that encodes information about both the physical conditions and the spatial location.

This shared representation is then passed to two independent output heads:

1. **Monotonic Head:** A dedicated MLP that takes the shared features and predicts the three monotonic variables: density ($\rho$), velocity ($v_1$), and temperature ($T$). This head is specifically constrained by physics-based loss terms that enforce the expected monotonic behavior for each variable (non-decreasing for $\rho$ and $T$, non-increasing for $v_1$).

2. **Non-Monotonic Head:** A second, parallel MLP that also takes the shared features but is tasked with predicting the two non-monotonic variables: the pressure difference ($p_{22} - p$) and the heat flux ($-q_1$). This head is not subjected to monotonicity constraints, allowing it to freely learn the characteristic peaks of these variables.

The outputs of the two heads are concatenated and passed through a final dense layer and a sigmoid activation function to produce the five final, normalized predictions. To enable uncertainty quantification, Monte Carlo Dropout is implemented by adding dropout layers within both the core networks and the output heads, which remain active during inference to generate a distribution of predictions.

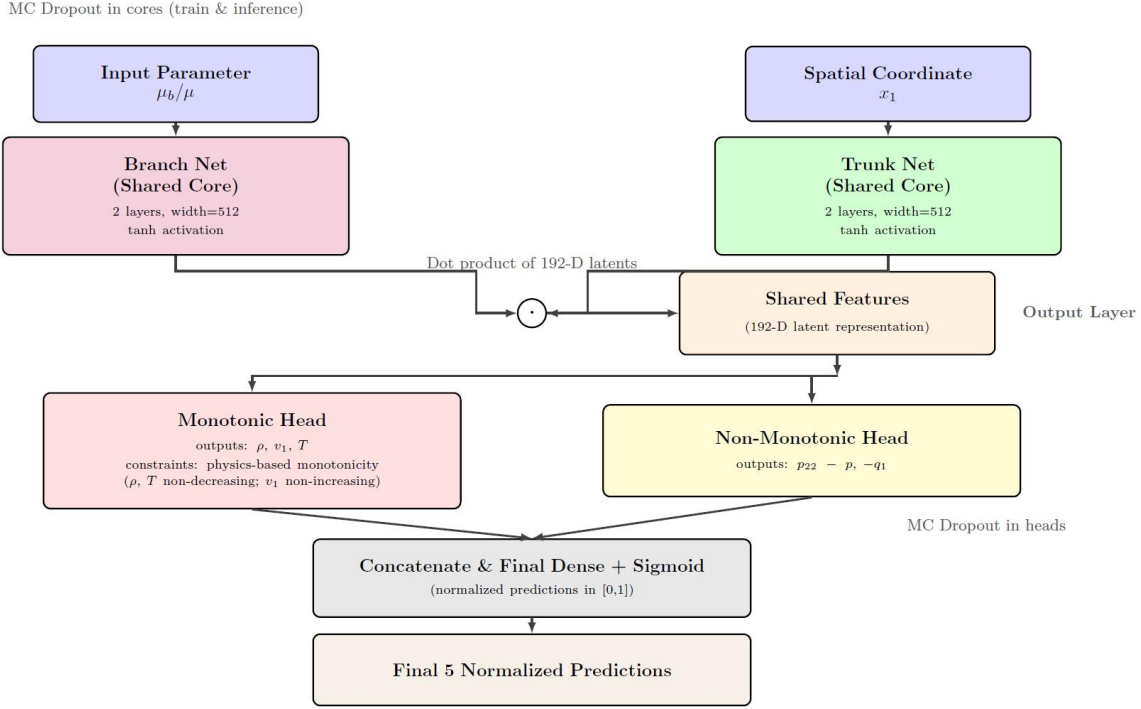

*Fig. 8: Multi-head physics-informed DeepONet (Section 5.4). A shared DeepONet core (Branch: layers, width=; Trunk: layers, width=; tanh activations) maps ($\mu_b/\mu$, $x_1$) to two -D latent vectors, combined by a dot product to form shared features. These feed two heads: a monotonic head for ($\rho, v_1, T$) with physics-based monotonicity constraints, and a non-monotonic head for ($p_{22} - p, -q_1$) (negative sign for $q_1$ chosen for loss consistency). The head outputs are concatenated and passed through a final dense layer with a sigmoid to produce five normalized predictions in [0,1]. MC dropout (active during both training and inference) is applied in both cores and heads for uncertainty quantification.*

# Results and Discussions

# Without Physical Constraints

Figure 9 shows the learning curves for the shock–wave surrogate trained *only* with the data (MSE) loss. The physical constraints are *not* enforced in the objective—only monitored as the dotted "total physics loss" trace—so the optimizer does not act on those residuals. Accordingly, the training/validation losses decay rapidly and then flatten, while the monitored physics residual remains small but exhibits non-monotonic wiggles due to mini–batch reshuffling and adaptive steps; these oscillations are expected when the physics terms are not weighted in the loss. The small gap between training and validation suggests limited overfitting, yet the absence of constraints explains the mild profile oscillations and local non-physical behavior seen in the unconstrained predictions. This plot serves as the *baseline* for comparison with the physics-enforced model, where adding constraint terms yields smoother, more physically consistent convergence.

Figures 10 and 11 present the predictions of the DeepONet model trained without any physical constraints for two unseen test cases with viscosity ratio $\mu_b/\mu = 50$. The results for $\mu_b/\mu = 500$ show a similar behavior and are not shown here. In each frame, the mean predictions (dashed red lines) are compared with the true simulation data (solid blue lines), together with the associated uncertainty bounds ($\pm 2\sigma$). The network is able to reproduce the general shape of the density $\rho$, velocity $v_1$, stress difference $p_{22} - p$, heat flux $-q_1$, and temperature $T$. However, noticeable deviations are observed near the sharp gradients around $x_1 = 0$, where the absolute errors grow by several orders of magnitude. The absence of monotonicity enforcement leads to oscillatory behavior and bias in the predicted profiles, particularly visible in the stress and heat flux. We use, $R^2$ (coefficient of determination), $R^2 = 1 - \sum_i (y_i - \hat{y}_i)^2 / \sum_i (y_i - \bar{y})^2$ to quantify the fraction of variance in the target explained by the model. Values near 1 indicate a good statistical fit, but they do *not* guarantee physical fidelity. Although several $R^2$ scores are moderately high, the predictions are not physically reliable when no physics is enforced—the network can fit data while violating conservation or other structural laws. The final $R^2$ values are: density $= 0.981858$, $v_1 = 0.993828$, temperature $= 0.992179$, $p_{22} - p = 0.949792$, and $-q_1 = 0.925542$. This underscores the need to incorporate physical constraints for trustworthy predictions. In summary, these results highlight the limitations of training without physical constraints: although the model can fit the general trend of the solution, the absence of monotonicity and boundary enforcement leads to oscillatory, biased, and nonphysical predictions with substantially larger errors.

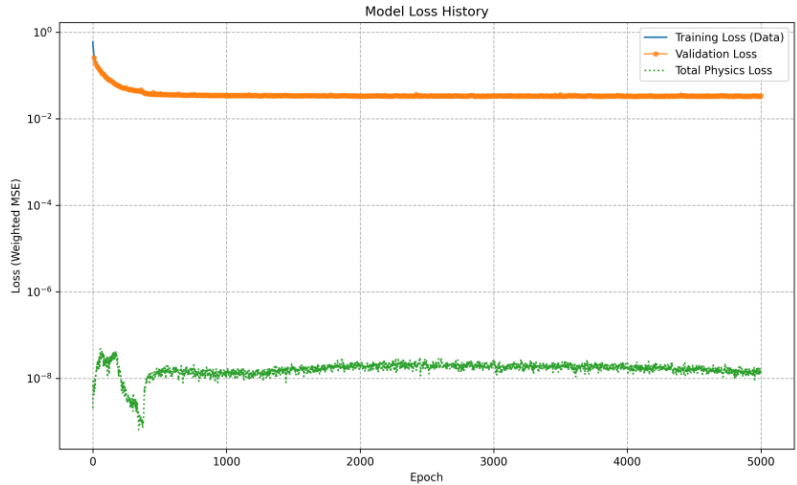

*Fig. 9: Training and validation loss history over epochs.*

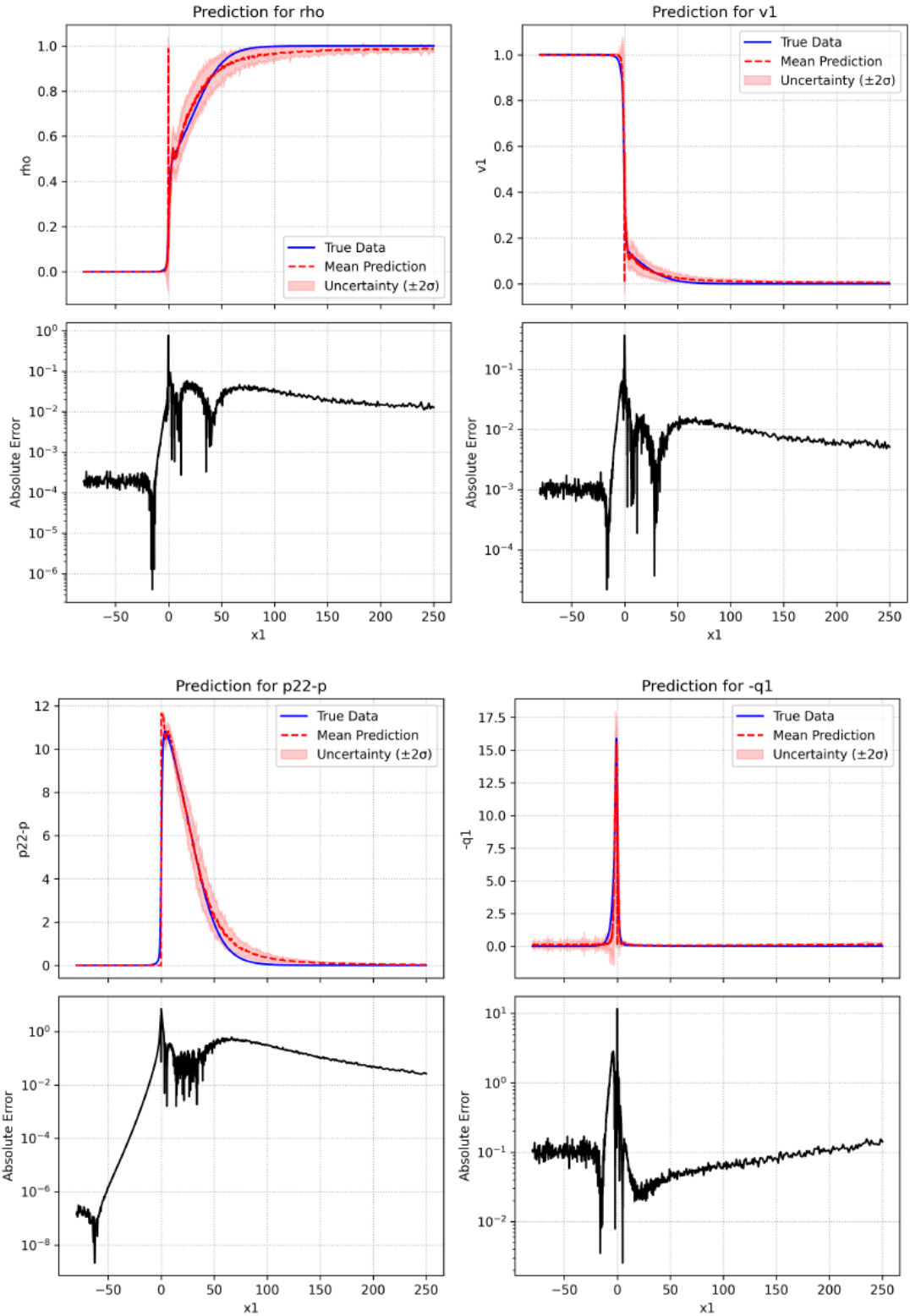

Fig. 10: Comparison of the DeepONet prediction without physical constraint (dashed red line) against the true simulation data (solid blue line) for the unseen test case where $\mu_b/\mu = 50$.

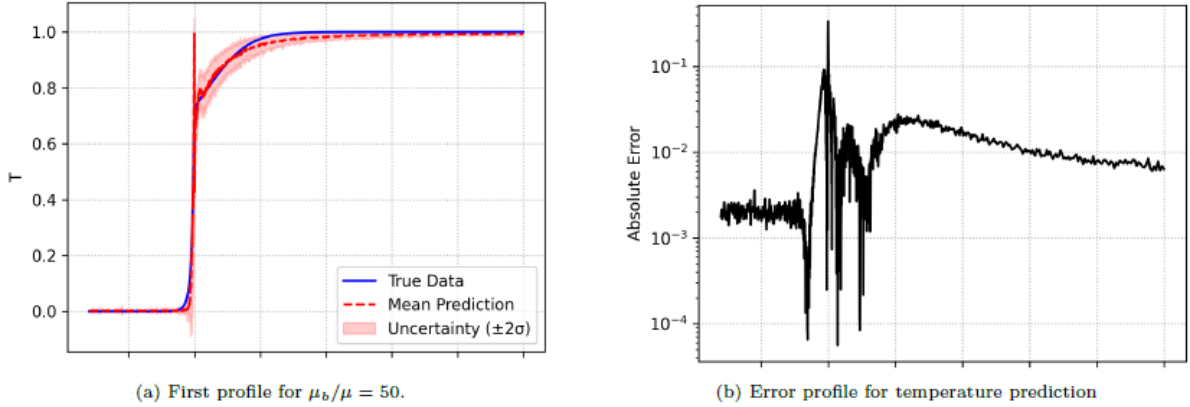

*Fig. 11: Comparison of two temperature profiles for the test case with $\mu_b/\mu = 50$.*

**Generalization to unseen viscosity ratios with physics-informed DeepONet.** With physics constraints enabled, the model delivers high-fidelity predictions for the two test ratios, $\mu_b/\mu = 50$ and 500. Figure 12 reports the training/validation loss history for the constrained model, showing a smooth, stable decay that is consistent with robust optimization under the augmented (data+physics) objective. The operator-learning design of DeepONet (branch for the parametric input $\mu_b/\mu$, trunk for the spatial basis over $x_1$) together with monotonicity and far-field constraints focuses the search on physically admissible solutions and promotes out-of-batch generalization.

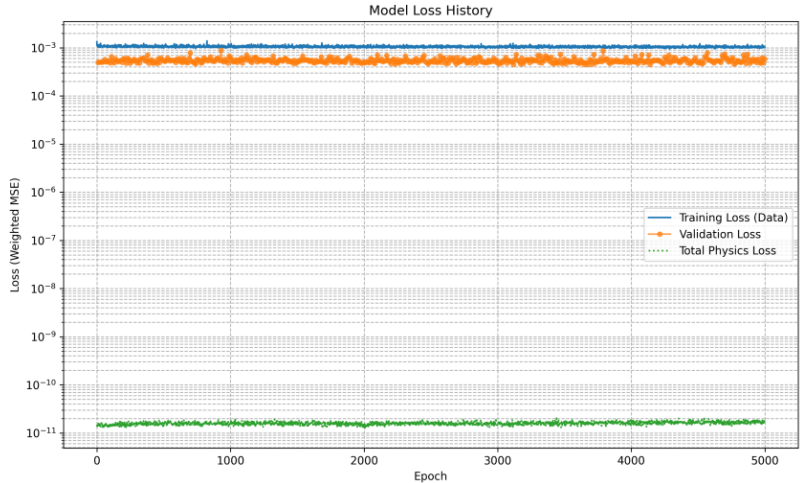

*Fig. 12: Training and validation loss history over epochs.*

Figure 13 plots the learning curves for the *physics–enforced* model: the data losses (training and validation MSE) and the aggregated physics loss used in the objective. After a brief transient, both data losses decrease steadily and remain closely aligned, indicating stable convergence with negligible generalization gap and no signs of overfitting. Simultaneously, the physics loss decays by orders of magnitude and settles at a low plateau, showing that the optimizer actively drives the monotonicity and far-field constraints toward satisfaction. Compared with the unconstrained baseline (cf. Fig. 9), the curves are smoother and largely monotone, with minimal oscillations—evidence that including the physical terms in the objective regularizes the training and yields a well-balanced optimization.

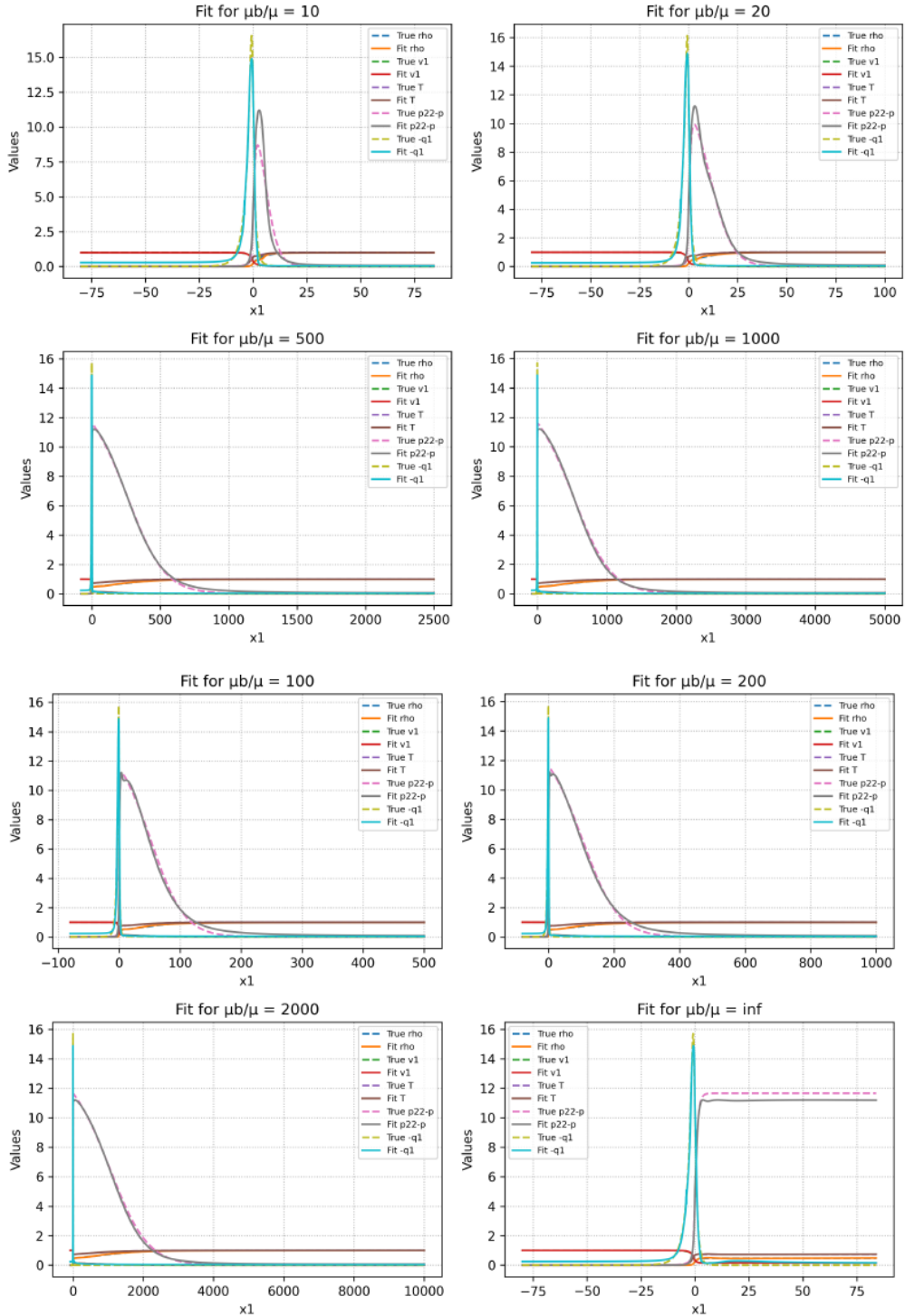

*Figure 13: Results without using the physical constraints.*

*Unseen case $\mu_b/\mu = 50$.* Figs. 14 and 15 visualize the predictions for all conserved/transport variables and a temperature zoom, respectively. The across-shock trends are captured cleanly (decreasing $\rho$ and $T$, increasing $v_1$), and the sharp layer is resolved without spurious oscillations; the temperature comparison in Fig. 15 highlights the agreement at the steepest gradient. Quantitatively, the constrained model attains near-perfect $R^2$ scores: $\rho = 0.999073$, $v_1 = 0.999726$, $T = 0.999393$, $p_{22} - p = 0.997613$, and $-q_1 = 0.994374$, confirming extremely small residual variance.

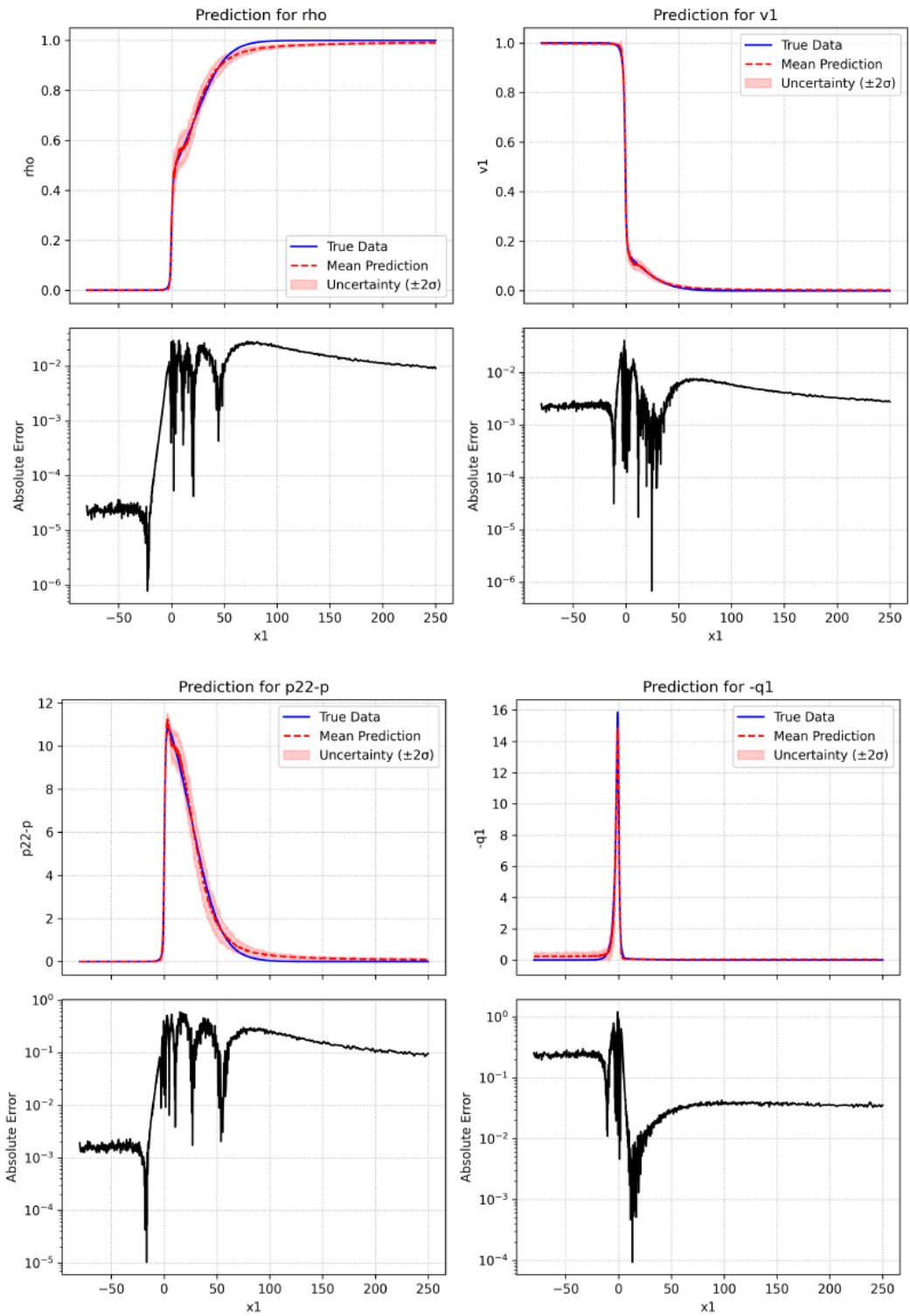

Figure 14: Comparison of flow properties profiles for the test case with $\mu_b/\mu = 50$.

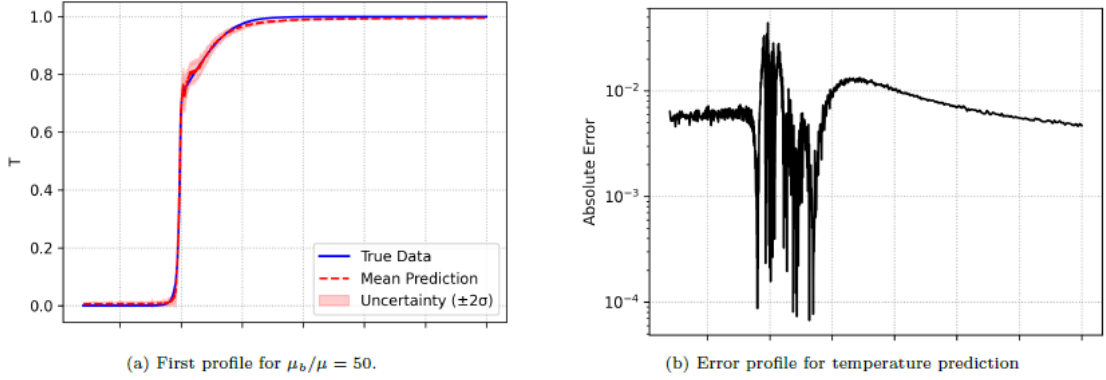

(a) First profile for $\mu_b/\mu = 50$.  (b) Error profile for temperature prediction

*Figure 15: Comparison of two temperature profiles for the test case with $\mu_b/\mu = 50$.*

*Unseen case $\mu_b/\mu = 500$.* Figs. 16 and 17 present the corresponding profiles and a temperature zoom for the harder case; the solution remains physically consistent, with smooth monotone jumps and accurate recovery of the stress and heat-flux surrogates. The $R^2$ values are likewise excellent: $\rho = 0.998092$, $v_1 = 0.999636$, $T = 0.999146$, $p_{22} - p = 0.998692$, and $-q_1 = 0.996112$.

Because $R^2$ measures the fraction of target variance explained by the model, values near one indicate an excellent statistical fit; here, unlike the unconstrained baseline, the high $R^2$ coincides with physically reliable solutions since the loss enforces monotonicity and far-field behavior.[1]

Overall, these figures corroborate that the physics-enforced DeepONet recovers the shock structure with minimal residual error and without nonphysical oscillations; the synergy between operator learning (branch/trunk), monotonicity and far-field constraints, and feature engineering yields accurate, physically consistent generalization to new viscosity ratios. :contentReference[oaicite:7]index=7

We reported the coefficient of determination, as introduced earlier, which measures the fraction of variance in the target explained by the model; $R^2 \approx 1$ indicates an excellent statistical fit. Because physics constraints are enforced in the loss, high $R^2$ here coincides with physically reliable solutions (unlike the unconstrained baseline).

All quantities exhibit $R^2 \geq 0.994$ in the harder case and $R^2 \geq 0.997$ for the stress component at $\mu_b/\mu = 50$, confirming that the network recovers the shock structure with extremely small residual variance. The slightly lower $R^2$ for the heat-flux surrogate ($-q_1$) is expected due to its higher sensitivity to gradients, but values $\geq 0.994$ still reflect excellent agreement. Taken together, these outcomes show that the integrated design—DeepONet decomposition, monotonicity and far-field constraints, and feature engineering—enables the model to learn the underlying physical *operator*, yielding accurate and physically consistent predictions at new parametric conditions.

---

[1] $R^2 = 1 - \sum_i (y_i - \hat{y}_i)^2 / \sum_i (y_i - \bar{y})^2$; high values alone do *not* guarantee physical fidelity unless the training objective encodes the physics.

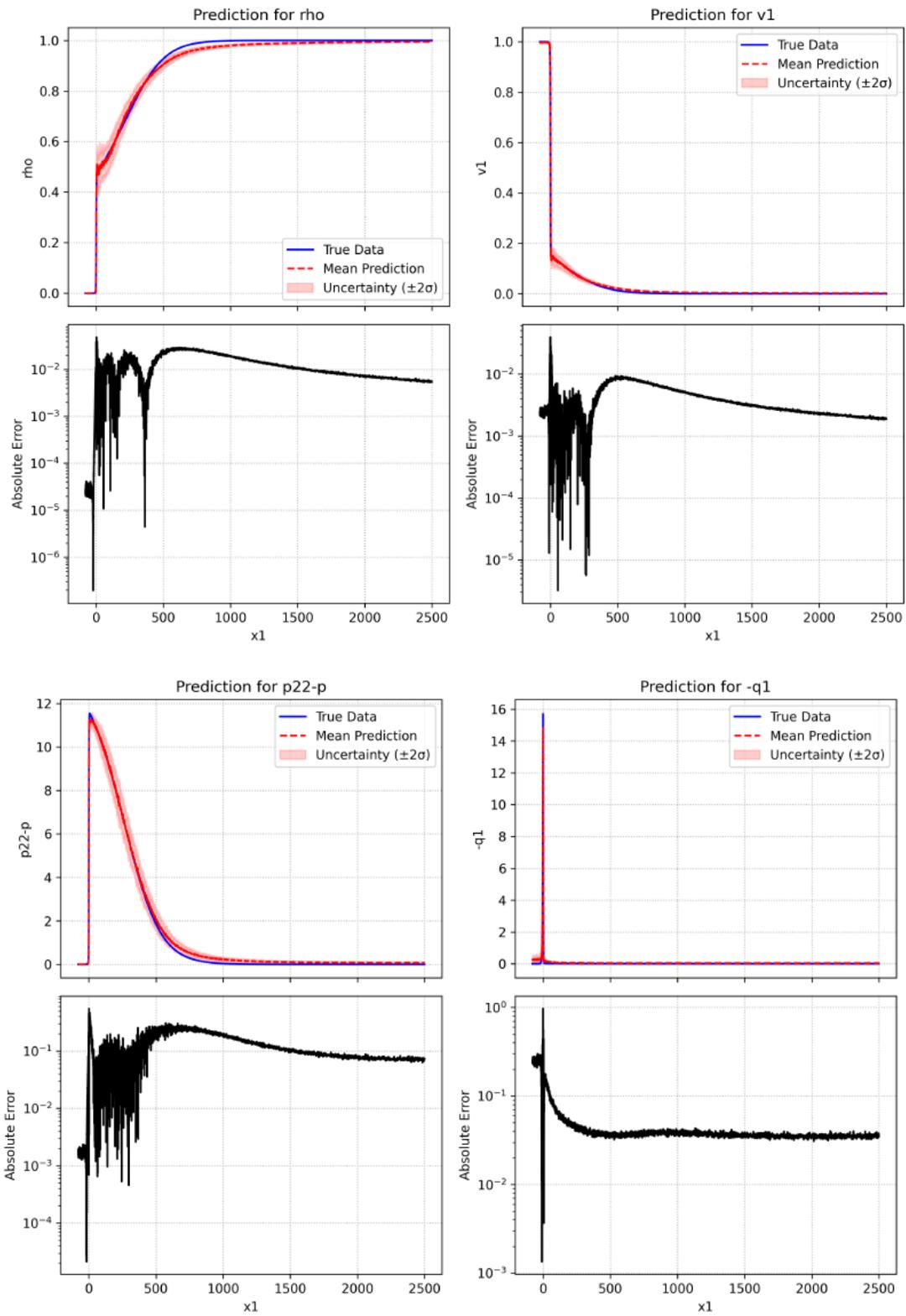

Figure 16: Comparison of flow properties profiles for the test case with $\mu_b/\mu = 500$.

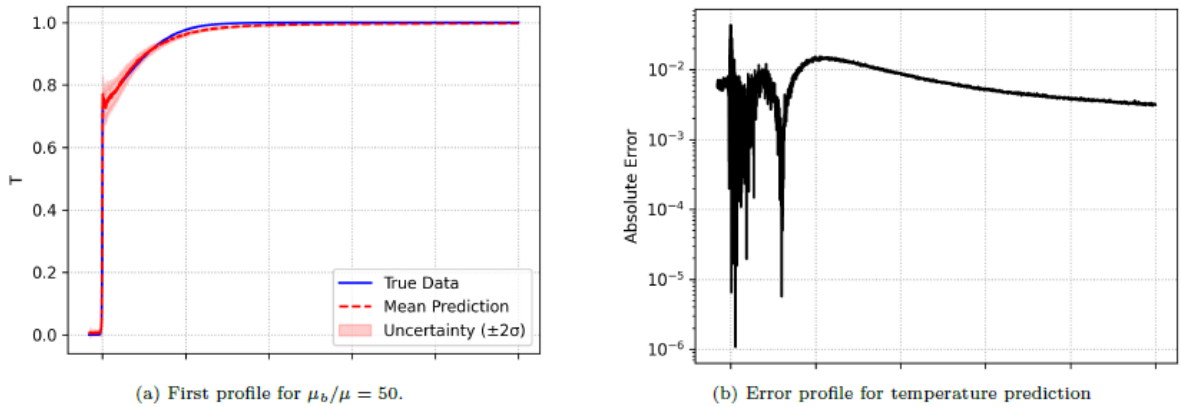

*Fi. 17: Comparison of two temperature profiles for the test case with $\mu_b/\mu = 500$.*

# Neural Network for Rarefied Hypersonic Flow over a Cylinder

To address the challenge of predicting hypersonic flow over a cylinder, a Deep Operator Network (DeepONet) was developed. This section details the model architecture, data handling, and training strategy employed to create a robust surrogate model capable of both interpolation and extrapolation. The rarefied hypersonic cylinder flow was the target of studies of previous research using the DSMC technique (Lofthouse et al. 2007; Bird 2013; B. Goshayeshi et al. 2015; Bijan Goshayeshi et al. 2015)

## Model Architecture and Hyperparameters

The core of the model is an ensemble of five DeepONets, a design chosen to enhance robustness and enable uncertainty quantification. Each DeepONet consists of two main sub-networks, a Branch Net and a Trunk Net, with increased capacity for improved accuracy, as depicted in Figure 18.

- **Branch Network:** This network processes the input physical parameter, which is the Mach number of the free-stream flow. It consists of four dense layers with 256 neurons each, using the hyperbolic tangent (tanh) activation function.

- **Trunk Network:** This network processes the spatial coordinates $(x, y)$ of the flow field. It mirrors the Branch Net's architecture with four dense layers of 256 neurons and tanh activation.

The outputs of these two networks are combined via a dot product, and the result is passed through a final dense layer to produce predictions for the two target variables: Mach number (MA), Temperature (TOV), and Pressure (P).

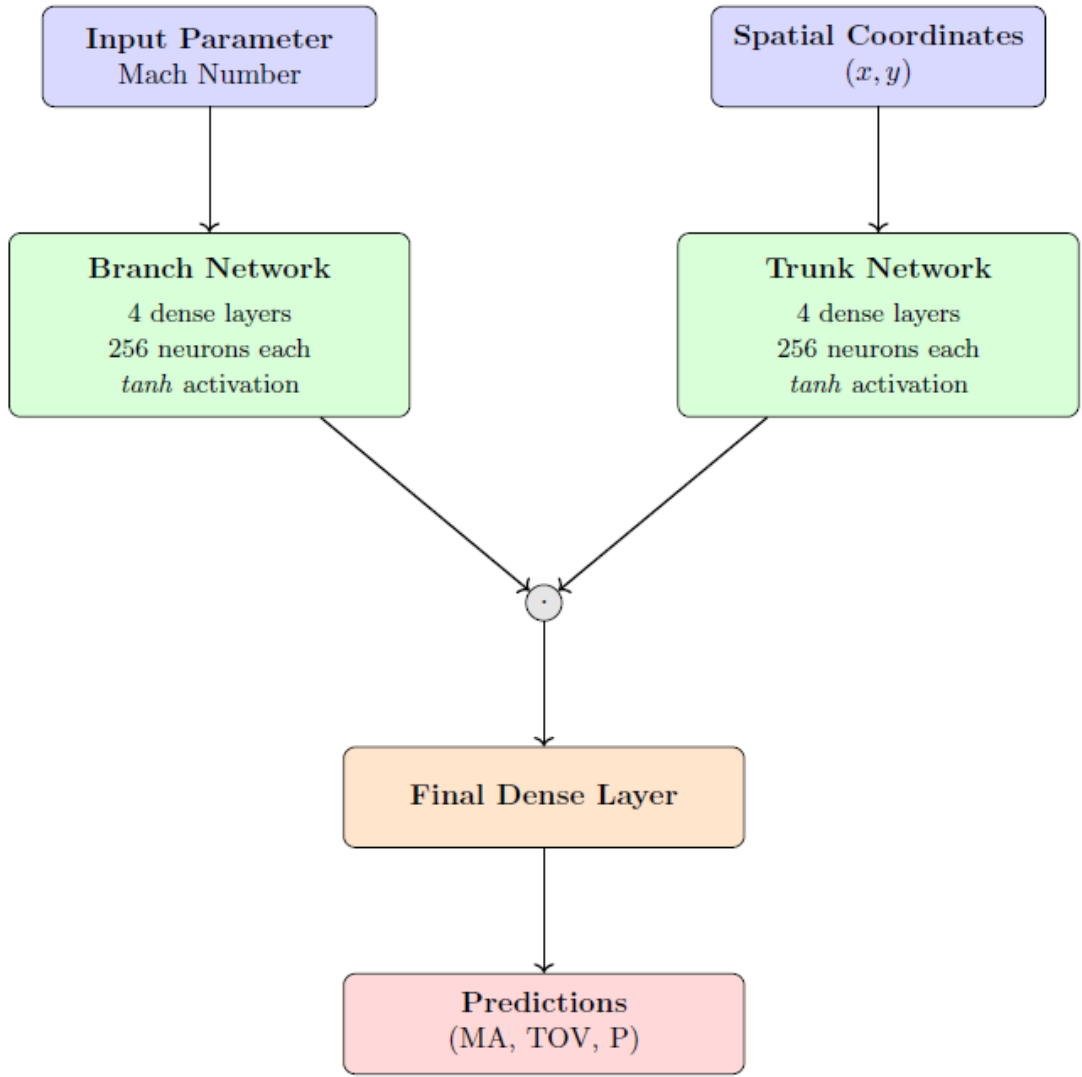

*Fig. 18: Schematic of the DeepONet architecture for the hypersonic cylinder problem. It features a Branch Net for the physical parameter (Mach number) and a Trunk Net for spatial coordinates $(x, y)$. The outputs are combined via a dot product, processed by a final dense layer, and produce three predictions: Mach number (MA), Temperature (TOV), and Pressure (P).*

## Data and Preprocessing

The model was trained on a sparse dataset derived from direct numerical simulations. The training set consists of simulations for hypersonic flow over a cylinder at Mach numbers M = 5, 7, and 9. The model was then evaluated on a comprehensive test set including both interpolation cases (M = 5.5, 6, 6.5, 7.5, 8, 8.5) and extrapolation cases (M = 9.5, 10).

Before training, all input and output data were scaled. The spatial coordinates $(x, y)$ and the input Mach number were scaled to a [0, 1] range using a 'MinMaxScaler'. The target output variables (MA, TOV, P) were standardized using a 'StandardScaler' to have a zero mean and unit variance, which stabilizes the training process.

## Custom Weighted Loss and Training

To improve the model's accuracy, particularly for the pressure field which is often the most challenging to predict, a custom weighted loss function was implemented. This function applies a higher penalty to errors in the pressure prediction compared to the other variables. Specifically, the Mean Squared Error (MSE) for pressure was weighted five times more heavily than the errors for Mach number and temperature.

The ensemble of five models was trained for 500 epochs using the Adam optimizer with a learning rate of $5 \times 10^{-4}$. To prevent overfitting and guide the training, two callbacks were used:

- **Early Stopping:** Training is halted if the validation loss does not improve for 100 consecutive epochs.

- **ReduceLROnPlateau:** The learning rate is automatically reduced if the validation loss plateaus for 40 epochs.

This setup ensures that each model in the ensemble is trained until it reaches its optimal performance on the validation set. The final prediction is the mean of the ensemble's outputs, and the standard deviation is used to quantify the model's uncertainty.

**Cross–Mach trends.** 33 visualizes what the network has learned across the parametric dimension by sweeping the free-stream Mach number from $M = 5$ to $M = 10$ and overlaying the resulting contours for the principal flow quantities (Mach number, temperature, and pressure) from the DSMC simulations. The figure exposes coherent, physically consistent trends: as $M$ increases, the detached bow shock strengthens and its stand-off distance decreases; the post-shock plateaus in temperature and pressure rise and contract toward the body; and the high-gradient region near the stagnation line becomes sharper. The wake likewise reorganizes with $M$, showing a systematic downstream displacement of iso-levels as the shock angle steepens. Taken together, these cross-Mach envelopes indicate that the model has not merely memorized individual fields; it has learned an operator $M \mapsto \{MA, T, P\}(x_1, x_2)$ that transports the loci of strong gradients (shock and near-stagnation layers) and rescales the post-shock states in a manner consistent with compressible-flow theory. This parametric consistency—visible simultaneously in all three variables—provides qualitative evidence for the model's ability to generalize within the training range and to capture the coupled dependence of the shock structure on $M$.

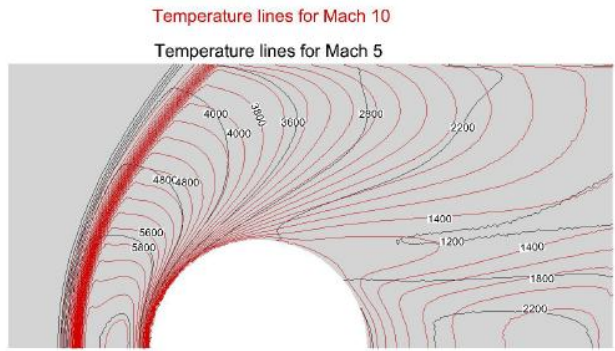

(a) Temperature Isolines

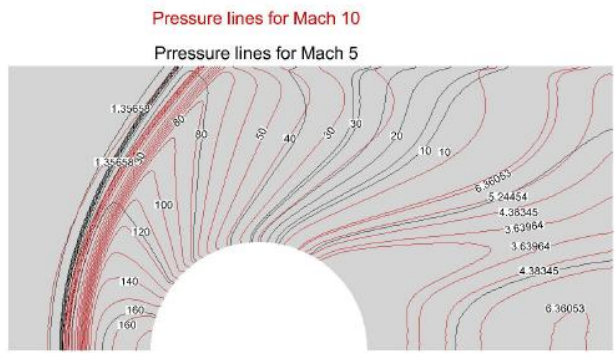

(b) Pressure Isolines

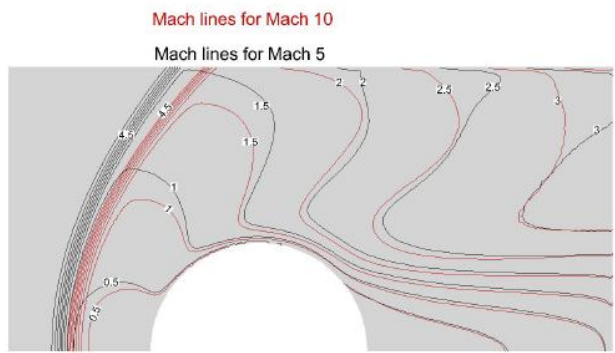

(c) Mach Number Isolines

*Fig. 19: Cross-Mach Trend from Mach 5 to 10*

Figure 20 shows the training (solid blue) and validation (dashed blue) loss curves for the cylinder test case on a logarithmic scale. Both losses drop sharply—by more than three orders of magnitude—during the first few epochs, then decrease in a step-like manner consistent with scheduled learning-rate decays. The final losses are small and nearly identical (training $\approx 1.7 \times 10^{-5}$, validation $\approx 1.5 \times 10^{-5}$), indicating a negligible generalization gap. Mild early oscillations reflect stochastic mini-batch effects, while the late-epoch plateau suggests convergence without overfitting.

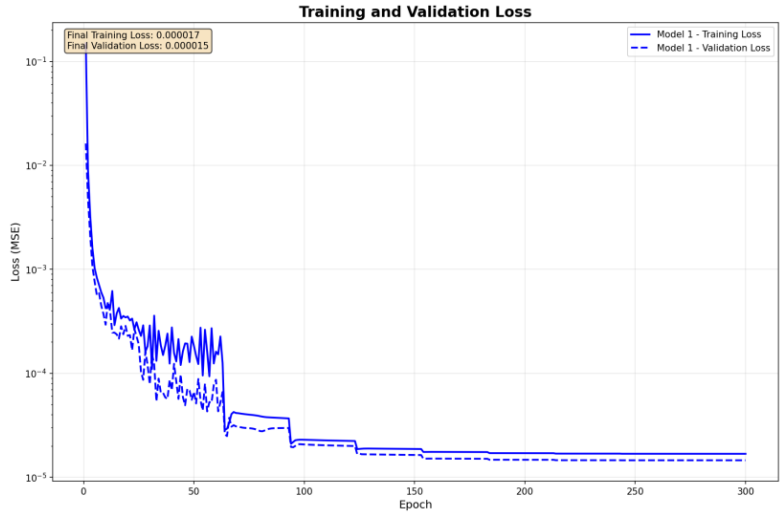

*Fig. 20: Training and validation loss curve.*

Figure 21 shows *overlaid* contour lines comparing the neural–operator prediction with the DSMC reference for the unseen case $M = 5.5$. Rather than side-by-side panels, both solutions are plotted on the same axes for each field (Mach number, pressure, temperature), so agreement appears as coincident iso-lines across the domain. The bow-shock shape, stagnation region, and downstream recovery are captured with near-indistinguishable contours; small, localized offsets are visible only across the steep shock layer, while the boundary-layer and far-field iso-line spacing is essentially identical. Overall, the overlay confirms high topological fidelity (shock location and curvature) and quantitative consistency of the NN with DSMC at this intermediate Mach number.

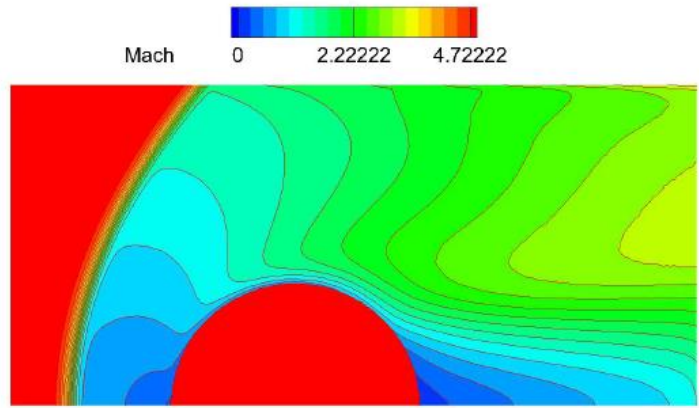

(a) Mach number distribution for the Mach 5.5 case.

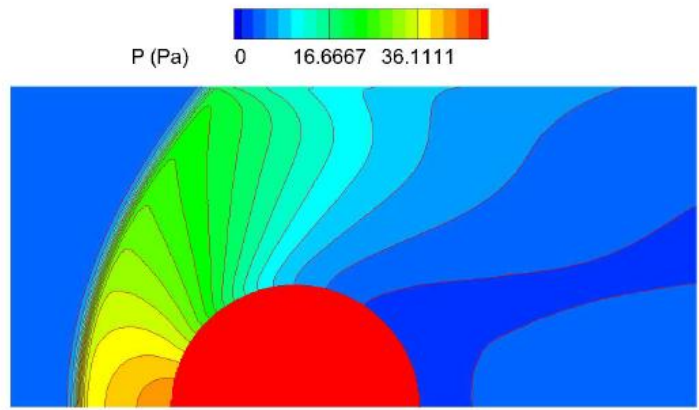

(b) Pressure distribution for the Mach 5.5 case.

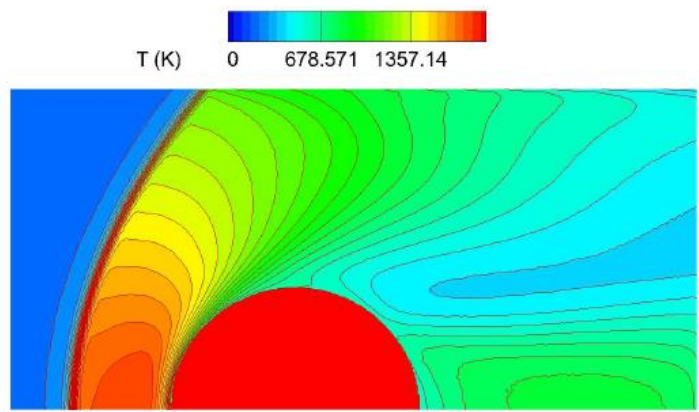

(c) Temperature distribution for the Mach 5.5 case.

*Fig. 21: Predicted flow fields for the hypersonic cylinder flow at M=5.5.*

## Uncertainty Quantification for M=10 Extrapolation

Uncertainty quantification was performed using a deep ensemble method. The standard deviation of the predictions from the five ensemble models serves as the uncertainty metric. Figure 22 shows the spatial distribution of this uncertainty for the Mach 10 extrapolation case, which is outside the training range. The figure shows two distinct regions, i.e., Low Uncertainty (Blue Regions): In these areas, all models in the ensemble are in strong agreement, indicating a high degree of confidence in the prediction.High Uncertainty (Red/Yellow Regions): In these areas, there is significant disagreement among the models, signaling low confidence. As expected, the highest uncertainty is concentrated in regions of high physical gradients, specifically the oblique shock wave. This is because learning sharp discontinuities is challenging, and the model is extrapolating to an unseen condition. This analysis is crucial as it identifies precisely where the model's predictions are most reliable and where they should be treated with caution.

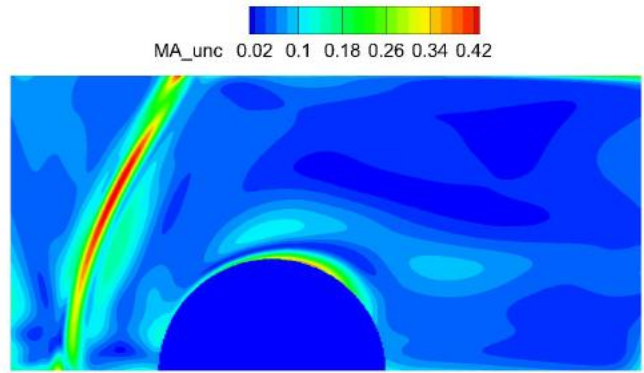

(a) Uncertainty (standard deviation) for Mach Number (MA) prediction.

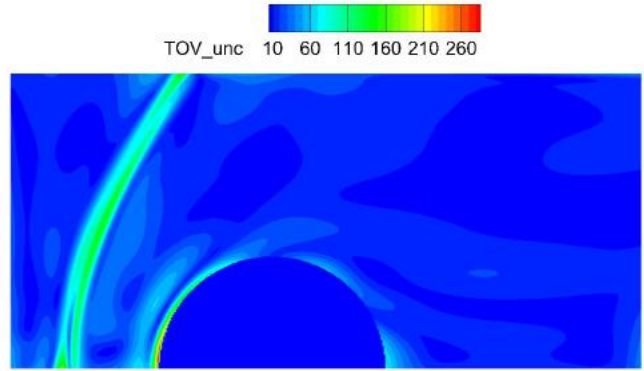

(b) Uncertainty (standard deviation) for Temperature (TOV) prediction.

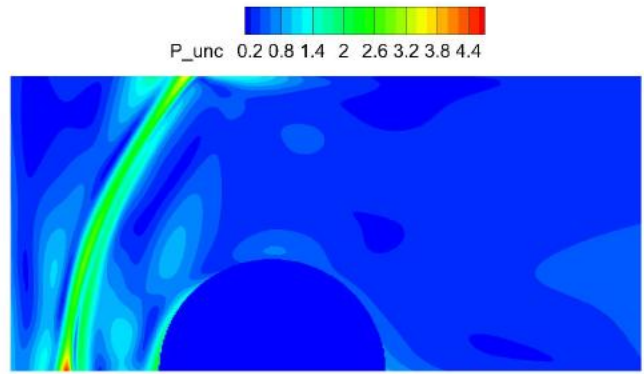

(c) Uncertainty (standard deviation) for Pressure (P) prediction.

*Fig. 22: Uncertainty (standard deviation) maps for the M=10 extrapolation case.*

# Quantitative Performance Summary

To assess the accuracy of the proposed DeepONet surrogate, two standard error measures were employed. The *Mean Absolute Error (MAE)* is defined as the average absolute deviation between the predicted values $\hat{y}_i$ and the reference data $y_i$, providing a scale-dependent indicator of accuracy. The *Mean Absolute Percentage Error (MAPE)*, on the other hand, expresses the average relative error in percentage form, thereby allowing for variable-to-variable comparisons independent of magnitude.

The quantitative results of these metrics are summarized in Tables [tab:mae_summary] and [tab:mape_summary]. The results demonstrate the model's robust generalization capability, as the error levels remain stable across both interpolation and extrapolation cases. This stability indicates that the network has successfully learned the underlying physical operator rather than simply memorizing the training data. For example, the MAPE for pressure is nearly the same for an interpolation case at $M = 6.5$ and an extrapolation case at $M = 10.0$, highlighting the reliability of the surrogate across regimes.

A closer inspection of the variables shows that the model predicted Temperature (TOV) with the highest accuracy, yielding an error of about 4% MAPE, followed by the Mach number (MA) with approximately 8% MAPE. Pressure (P) proved to be the most challenging variable, with a relative error of roughly 14.6%. Taken together, these findings confirm that the DeepONet framework is capable of providing accurate predictions of hypersonic cylinder flows even for test conditions that lie significantly outside the sparse training set.

Table 1: Mean Absolute Error (MAE) Summary

| File | Mach | Type | Gap | MAE (MA) | MAE (TOV) | MAE (P) |
|---|---|---|---|---|---|---|
| GridM=5.5 | 5.5000 | Interpolation | 0.5000 | 0.0647 | 22.5679 | 0.4442 |
| GridM=6 | 6.0000 | Interpolation | 1.0000 | 0.0645 | 22.4905 | 0.4436 |
| GridM=6.5 | 6.5000 | Interpolation | 0.5000 | 0.0644 | 22.4283 | 0.4433 |
| GridM=7.5 | 7.5000 | Interpolation | 0.5000 | 0.0643 | 22.3344 | 0.4427 |
| GridM=8 | 8.0000 | Interpolation | 1.0000 | 0.0643 | 22.2991 | 0.4424 |
| GridM=8.5 | 8.5000 | Interpolation | 0.5000 | 0.0644 | 22.2706 | 0.4421 |
| GridM=9.5 | 9.5000 | Extrapolation | 0.5000 | 0.0647 | 22.2340 | 0.4428 |
| GridM=10 | 10.0000 | Extrapolation | 1.0000 | 0.0651 | 22.2269 | 0.4444 |

Table 2: Mean Absolute Percentage Error (MAPE) Summary

| File | Mach | Type | Gap | MAPE (MA) | MAPE (TOV) | MAPE (P) |
|---|---|---|---|---|---|---|
| GridM=5.5 | 5.50 | Interpolation | 0.50 | 8.11% | 4.01% | 14.73% |
| GridM=6 | 6.00 | Interpolation | 1.00 | 8.07% | 4.00% | 14.66% |
| GridM=6.5 | 6.50 | Interpolation | 0.50 | 8.04% | 3.98% | 14.61% |
| GridM=7.5 | 7.50 | Interpolation | 0.50 | 8.04% | 3.96% | 14.55% |
| GridM=8 | 8.00 | Interpolation | 1.00 | 8.06% | 3.96% | 14.53% |
| GridM=8.5 | 8.50 | Interpolation | 0.50 | 8.10% | 3.95% | 14.52% |
| GridM=9.5 | 9.50 | Extrapolation | 0.50 | 8.21% | 3.95% | 14.61% |
| GridM=10 | 10.00 | Extrapolation | 1.00 | 8.29% | 3.95% | 14.75% |

# Conclusions

This work introduced physics–influenced neural surrogates for rarefied gas dynamics spanning three tiers of difficulty: (i) BGK kinetic relaxation, (ii) the one–dimensional structure of a polyatomic standing shock, and (iii) two–dimensional hypersonic flow over a cylinder learned as a parametric operator. By embedding carefully designed physical constraints into PINN/DeepONet training, we demonstrated that neural operators can be made accurate, data–efficient, and physically trustworthy across regimes where classical continuum closures break down. For the BGK relaxation problem, we reformulated the learning target as the perturbation from Maxwell–Boltzmann equilibrium, $P(v,t) = P_M(v)\,[1 + \Phi(v,t)]$, which converts the BGK equation into a stable decay law for $\Phi$. This perturbation ansatz prevents collapse to nonphysical solutions, yields smooth convergence to equilibrium in the forward setting, and clarifies why the inverse problem (inferring $\tau$) is ill–posed without additional priors.

For rarefied Polyatomic shock (DeepONet with constraints) in the pseudo–$CO_2$, we used a physics–enforced DeepONet that couples a branch net (parameter: $\mu_b/\mu$) with a trunk net (space: $x_1$), and augments data loss with monotonicity and far–field gradient penalties. The model reproduces the non–equilibrium shock structure and generalizes to viscosity ratios not seen in training; notably, it predicts unseen cases at $\mu_b/\mu = 50$ and $500$ with high fidelity. For the Hypersonic cylinder problem, a DeepONet ensemble trained only at $M = 5,7,9$ accurately interpolates and extrapolates full flowfields over $M = 5.5$–$10$. A custom weighted loss improves pressure accuracy, while ensemble–based uncertainty quantification peaks where the flow has strong gradients, such as across the bow shock—providing actionable diagnostics for trust.

Across all testbeds, enforcing physics within neural operators (boundedness/monotonicity, far–field consistency, conservation–aware objectives) markedly reduces hypothesis space, stabilizes training, and elevates generalization—turning sparse, high–fidelity datasets into robust surrogates suitable for multi–query tasks (design sweeps, UQ, and inverse studies). Future directions include extending the kinetic surrogate to ES–BGK closures for tunable Pr effects, tightening conservation/entropy enforcement, scaling to 3D geometries and multi–physics (e.g., internal energy/vibrational nonequilibrium, chemical reactions), and integrating uncertainty–aware surrogates into end–to–end design and control loops.

# Acknowledgment

The authors would like to acknowledge Dr. E. Karniadakis for his hints on using DeepONet networks. They also thank Prof. S. Kosuge and K. Aoki for providing the data of polyatomic shock waves from their numerical simulations. The authors also acknowledge Dr. Nam T. P. Le from the Institute of Engineering and Technology, Thu Dau Mot University, Binh Duong Province, Vietnam, for hints on establishing the neural network for the cylinder problem.

# References

Barwey, Shivam, Pinaki Pal, Saumil Patel, et al. 2025. "Mesh-Based Super-Resolution of Fluid Flows with Multiscale Graph Neural Networks." *Computer Methods in Applied Mechanics and Engineering* 443: 118072.

Bird, G. A. 2013. *The DSMC Method*. CreateSpace Independent Publishing Platform.

Bird, Graeme A. 1994. *Molecular Gas Dynamics and the Direct Simulation of Gas Flows*. Clarendon Press.

Cai, Sheng, Zhiping Mao, Zhiping Wang, Ming Yin, and George Em Karniadakis. 2021. "Physics-Informed Neural Networks for Fluid-Structure Interactions." *Journal of Computational Physics* 446: 110612.

Chen, Xiaowei, Zhaohui Qu, Yanan Wang, et al. 2025. "Prediction of Spectral Response for Explosion Separation Based on DeepONet." *Aerospace* 12 (4): 310.

Corbetta, A., A. Gabbana, V. Gyrya, D. Livescu, J. Prins, and F. Toschi. 2023. "Toward Learning Lattice Boltzmann Collision Operators." *The European Physical Journal E* 46 (3): 10.

Fowler, Eric, Conor J McDevitt, and Subrata Roy. 2024. "Physics-Informed Neural Network Simulation of Thermal Cavity Flow." *Scientific Reports* 14 (1): 15203.

Gallis, Michail A, John Robert Torczynski, Steven J Plimpton, Daniel J Rader, and Timothy Koehler. 2014. *Direct Simulation Monte Carlo: The Quest for Speed.* Sandia National Lab.(SNL-NM), Albuquerque, NM (United States).

Goshayeshi, Bijan, Ehsan Roohi, and Stefan Stefanov. 2015. "A Novel Simplified Bernoulli Trials Collision Scheme in the Direct Simulation Monte Carlo with Intelligence over Particle Distances." *Physics of Fluids* 27 (10).

Goshayeshi, B., E. Roohi, and S. Stefanov. 2015. "DSMC Simulation of Hypersonic Flows Using an Improved SBT-TAS Technique." *J. Comput. Phys.* 303: 28–44.

Griffith, W. C., and A. Kenny. 1957. "On fully-dispersed shock waves in carbon dioxide." *Journal of Fluid Mechanics* 3 (3): 286–88.

Guo, Xiaoxiao, Wei Li, and Francesco Iorio. 2016. "Convolutional Neural Networks for Steady Flow Approximation." *Proceedings of the 22nd ACM SIGKDD International Conference on Knowledge Discovery and Data Mining*, 481–90.

Karniadakis, George E, Ioannis G Kevrekidis, Lu Lu, Paris Perdikaris, Sifan Wang, and Liu Yang. 2021. "Physics-Informed Machine Learning." *Nature Reviews Physics* 3 (6): 422–40.

Kosuge, Shingo, and Kazuo Aoki. 2018. "Shock-Wave Structure for a Polyatomic Gas with Large Bulk Viscosity." *Physical Review Fluids* 3 (2): 023401.

Liu, Z., C. Zhang, W. Zhu, and D. Huang. 2024. "A Physics-Informed Neural Network Based on the Boltzmann Equation with Multiple-Relaxation-Time Collision Operators." *Axioms* 13 (9): 588.

Lofthouse, A. J., I. D. Boyd, and J. M. Wright. 2007. "Effects of Continuum Breakdown on Hypersonic Aerothermodynamics." *Phys. Fluids* 19: 027105.

Lu, Lu, Pengzhan Jin, Guofei Pang, Zhongqiang Zhang, and George E Karniadakis. 2021. "DeepONet: Learning Operators for Identifying Nonlinear Continuous Systems." *Nature Machine Intelligence* 3 (9): 774–84.

McDevitt, Conor J, and Xian-Zhu Tang. 2024. "A Physics-Informed Deep Learning Description of Knudsen Layer Reactivity Reduction." *Physics of Plasmas* 31 (6).

Miller, S. T., N. V. Roberts, S. D. Bond, and E. C. Cyr. 2022. "Neural-Network Based Collision Operators for the Boltzmann Equation." *Journal of Computational Physics* 470: 111541.

Moore, SG, A Borner, AK Stagg, TP Koehler, JR Torczynski, and MA Gallis. 2019. "Direct Simulation Monte Carlo on Petaflop Supercomputers and Beyond." *Physics of Fluids* 31 (8).

Peyvan, Ali, Vivek Kumar, and George E Karniadakis. 2025. "Fusion-DeepONet: A Data-Efficient Neural Operator for Geometry-Dependent Hypersonic and Supersonic Flows." *arXiv Preprint arXiv:2501.01934*.

Peyvan, Ali, Vismay Oommen, Ameya D Jagtap, and George E Karniadakis. 2024. "Riemannonets: Interpretable neural operators for Riemann problems." *Computer Methods in Applied Mechanics and Engineering* 426: 116996.

Raissi, Maziar, Paris Perdikaris, and George E Karniadakis. 2019. "Physics-Informed Neural Networks: A Deep Learning Framework for Solving Forward and Inverse Problems Involving Nonlinear Partial Differential Equations." *Journal of Computational Physics* 378: 686–707.

Roohi, Ehsan, Hassan Akhlaghi, and Stefan Stefanov. 2025. *Advances in Direct Simulation Monte Carlo: From Micro-Scale to Rarefied Flow Phenomena*. Springer Nature Singapore. https://doi.org/10.1007/978-981-96-8200-3.

Roohi, Ehsan, and Ahmad Shoja-Sani. 2025. "Data-Driven Surrogate Modeling of DSMC Solutions Using Deep Neural Networks." *Aerospace Science and Technology*, ahead of print. https://doi.org/10.1016/j.ast.2025.110785.

Sun, Luning, Han Gao, Shaowu Pan, and Jian-Xun Wang. 2020. "Surrogate Modeling for Fluid Flows Based on Physics-Constrained Deep Learning." *Computer Methods in Applied Mechanics and Engineering* 361: 112732.

Takao, Ryuta, and Satoshi Ii. 2025. "Fine-Tuning Physics-Informed Neural Networks for Cavity Flows Using Coordinate Transformation." *arXiv Preprint arXiv:2508.01122*.

Taniguchi, Shigeru, Takayuki Arima, Tommaso Ruggeri, and Masaru Sugiyama. 2014. "Thermodynamic Theory of the Shock Wave Structure in a Rarefied Polyatomic Gas: Beyond the Bethe-Teller Theory." *Physical Review E* 89 (1): 013025.

Tatsios, G., A. K. Chinnappan, A. Kamal, et al. 2025. "A DSMC-CFD Coupling Method Using Surrogate Modelling for Low-Speed Rarefied Gas Flows." *Journal of Computational Physics* 520: 113500.

Thuerey, Nils, Kay Weißenow, Lukas Prantl, and Xiangyu Hu. 2020. "Deep Learning Methods for Reynolds-Averaged Navier–Stokes Simulations of Airfoil Flows." *AIAA Journal* 58 (1): 251–64.

Vincenti, W. G., and C. H. Kruger. 1967. *Introduction to Physical Gas Dynamics*. John Wiley; Sons.

Wang, Sifan, Han-Xin Wang, and Paris Perdikaris. 2021. "Learning the Solution Operator of Parametric Partial Differential Equations with Physics-Informed DeepONets." *Science Advances* 7 (40): eabi8605.

Xiao, T., and M. Frank. 2021. "Using Neural Networks to Accelerate the Solution of the Boltzmann Equation." *Journal of Computational Physics* 443: 110521.

Xiao, Tian. 2025. "Solving Continuum and Rarefied Flows Using Differentiable Programming." *arXiv Preprint arXiv:2501.13478*.

Zanardi, I, S Venturi, A Munafò, and M Panesi. 2023. "Towards efficient simulations of non-equilibrium chemistry in hypersonic flows: neural operator-enhanced 1-D shock simulations." *AIAA SCITECH 2023 Forum*, 1202.

Zel'dovich, Y. B., and Y. P. Raizer. 2002. *Physics of Shock Waves and High-Temperature Hydrodynamic Phenomena*. Courier Corporation.

Zheng, Jiachen, Hui Hu, Jichao Huang, Ben Zhao, and He Huang. 2025. "CF-DeepONet: Deep operator neural networks for solving compressible flows." *Aerospace Science and Technology*, 110329.